\documentclass[]{aastex631}
\usepackage{CJK}

\usepackage{ulem}
\usepackage{soul}
\setul{}{1pt}

\begin{document}
\begin{CJK*}{UTF8}{ipxm}

\title{A Necessary Condition for the Submergence of Proto-Neutron Star Magnetic Fields by Supernova Fallback}

\author[0000-0002-0700-2223]{Akihiro Inoue}
\affiliation{Department of Earth and Space Science, Graduate School of Science, Osaka University, Toyonaka, Osaka 560-0043, Japan}
\affiliation{Department of Earth Science and Astronomy, The University of Tokyo, Meguro, Tokyo 153-8902, Japan}
\email{inoue-a@g.ecc.u-tokyo.ac.jp}

\author[0000-0003-3882-3945]{Shinsuke Takasao}
\affiliation{Department of Earth and Space Science, Graduate School of Science, Osaka University, Toyonaka, Osaka 560-0043, Japan}
\affiliation{Humanities and Sciences/Museum Careers, Musashino Art University, Kodaira, Tokyo 187-8505, Japan}

\author[0000-0003-4299-8799]{Kazumi Kashiyama}
\affiliation{Astronomical Institute, Tohoku University, Sendai, Miyagi 980-8578, Japan}
\affiliation{Kavli Institute for the Physics and Mathematics of the Universe, The University of Tokyo, Kashiwa, Chiba 277-8583, Japan}

\author[0000-0003-0805-8234]{Yici Zhong}
\affiliation{TAPIR, Mailcode 350-17, California Institute of Technology, Pasadena, CA 91125, USA}
\affiliation{Walter Burke Institute for Theoretical Physics, California Institute of Technology, Pasadena, CA 91125, USA}

\author[0000-0003-0114-5378]{Hiroyuki R. Takahashi}
\affiliation{Department of Natural Sciences, Faculty of Arts and Sciences, Komazawa University, Setagaya, Tokyo 154-8525, Japan}

\begin{abstract}

Central compact objects (CCOs) are a subclass of neutron stars with a dipole magnetic field strength considerably weaker than those of radio pulsars and magnetars.
One possible explanation for such weak magnetic fields in the CCOs is the hidden magnetic field scenario, in which supernova fallback submerges the magnetosphere of a proto-neutron star beneath a newly formed crust.
However, the fallback mass and time scale required for this submergence process remain uncertain.
We perform one-dimensional general relativistic magnetohydrodynamic simulations of the supernova fallback onto a magnetized proto-neutron star, while considering neutrino cooling.
In our simulations, the infalling material compresses the magnetic field and drives a strong shock.
The shock initially expands outward, but eventually stalls and recedes as neutrino cooling becomes significant.
After the shock stalls, the gas density above the magnetosphere increases rapidly, potentially leading to the formation of a new crust.
To understand the shock dynamics, we develop semi-analytic models that describe the resulting magnetospheric and shock radii when the shock stalls.
By comparing the fallback time scale with the shock stalling time scale, corresponding to the waiting time for the new crust formation, we derive a necessary condition for the submergence of the proto-neutron star's magnetic field.
Our results will provide guidance for investigating the diversity of young isolated neutron stars through multidimensional simulations.

\end{abstract}



\section{Introduction} \label{sec:intro}

Based on multi-wavelength observations, young isolated neutron stars (NSs) in our galaxy with ages of $\lesssim1-10~{\rm kyr}$ are classified as magnetars, radio pulsars, and central compact objects (CCOs).
The central engines powering these objects are believed to be the magnetic energy, rotation energy, and thermal energy of the NSs, respectively \citep[e.g., see reviews by][]{Enoto2019,Borghese2023}.
Their characteristic values of the dipole magnetic field strength at the NS surface are considerably different: $>10^{14}~{\rm G}$ for magnetars, $10^{12-13}~{\rm G}$ for radio pulsars, and $<10^{11}~{\rm G}$ for CCOs.
However, the origin of such diversity remains unclear.

What determines the magnetic field strength of NSs is an open question \citep[e.g., see reviews by][]{Igoshev2021}.
When a proto-NS (PNS) forms in a core-collapse supernova, the magnetic field of the progenitor core is expected to undergo amplification through various mechanisms: magnetic flux conservation \citep{Woltjer1964}, turbulent dynamo action \citep[][]{Duncan1992,Thompson1993,Raynaud2020,Masada2022}, magnetorotational instability inside the PNS \citep[][]{Akiyama2003,Obergaulinger2009,Masada2015,Reboul2021}, and/or steady accretion shock instability of the post-bounce core-collapse supernova environment \citep{Blondin2003,Endeve2010,Endeve2012}.
On the other hand, it is thought that the NS magnetic field decays over a relatively long time scale, $\gg 10~{\rm kyr}$, due to Ohmic diffusion, ambipolr diffusion, and Hall drift \citep[e.g.,][]{Goldreich1992,Takahashi2011,Skiathas2024}.

Supernova fallback has also been proposed as a contributing factor in the magnetic field evolution, particularly in the formation of the CCOs \citep[][]{Bernal2010,Ho2011,Shabaltas2012}.
The fallback phase starts several seconds or more after the core bounce at the time when the neutrino-driven wind subsides \citep[see e.g.,][]{Thompson2001_wind,Janka2022,Wang2023,Wang2024}.
If the fallback accretion rate is sufficiently high, the PNS's magnetic field will be submerged beneath a crust newly formed by the fallback material \citep{Young1995,Muslimov1995,Geppert1999}.
This case corresponds to the CCO formation \citep[][]{Torres2016,Shigeyama2018}.
This scenario is referred to as the hidden magnetic field scenario.
Depending on the fallback mass and the magnetic field strength, the hidden magnetic fields are expected to re-emerge through the crust on a time scale of $10^{1-2}~{\rm kyr}$ or longer \cite[][]{Ho2011,Vigano2012,Igoshev2016,Gourgouliatos2020,Igoshev2021_a,Dehman2023,Fraija2025}.
The hidden magnetic field is also believed to be the energy source of giant flares observed in soft gamma-ray repeaters \citep[][]{Thompson2001}.

The fallback accretion rate remains highly uncertain.
Based on numerical simulations of the supernova explosion \citep[e.g.,][]{Zhang2008,Dexter2013,Moriya2018}, the fallback accretion rate is approximated to be \citep{Metzger2018,Barrere2022}
\begin{eqnarray}
    \dot{M}_{\rm fb}
    =
    \frac{2}{3} \frac{M_{\rm fb}}{t_{\rm fb}}
    \left(\frac{1}{1+t/t_{\rm fb}}\right)^{5/3},
    \label{eq:fallback}
\end{eqnarray}
where $t$ stands for time, and $M_{\rm fb}$ and $t_{\rm fb}$ are the typical fallback mass and fallback time scale, respectively.
The dependence of $\dot{M}_{\rm fb}\propto t^{-5/3}$ for $t\gg t_{\rm fb}$ is expected for an accretion of marginally gravitationally bound matter \citep{Michel1988,Chevalier1989}.
The typical fallback mass is estimated to be $M_{\rm fb}\sim 10^{-(4-1)}~{\rm M_\odot}$, depending on the progenitor structure \citep[see e.g.,][]{Ugliano2012,Ertl2016a,Ertl2016b}.
The fallback time $t_{\rm fb}$ typically ranges from $1~{\rm s}$ to $10^3~{\rm s}$ \citep{Janka2022}, depending on the mean density of the progenitor star layer from which the fallback mass originates \citep{Metzger2018}.
Therefore, $\dot{M}_{\rm fb}$ spans a wide range of $(10^{-7}-10^{-1})~{\rm M_\odot~s^{-1}}$.

\begin{figure*}[tb]
\centering
\includegraphics[width=\linewidth]{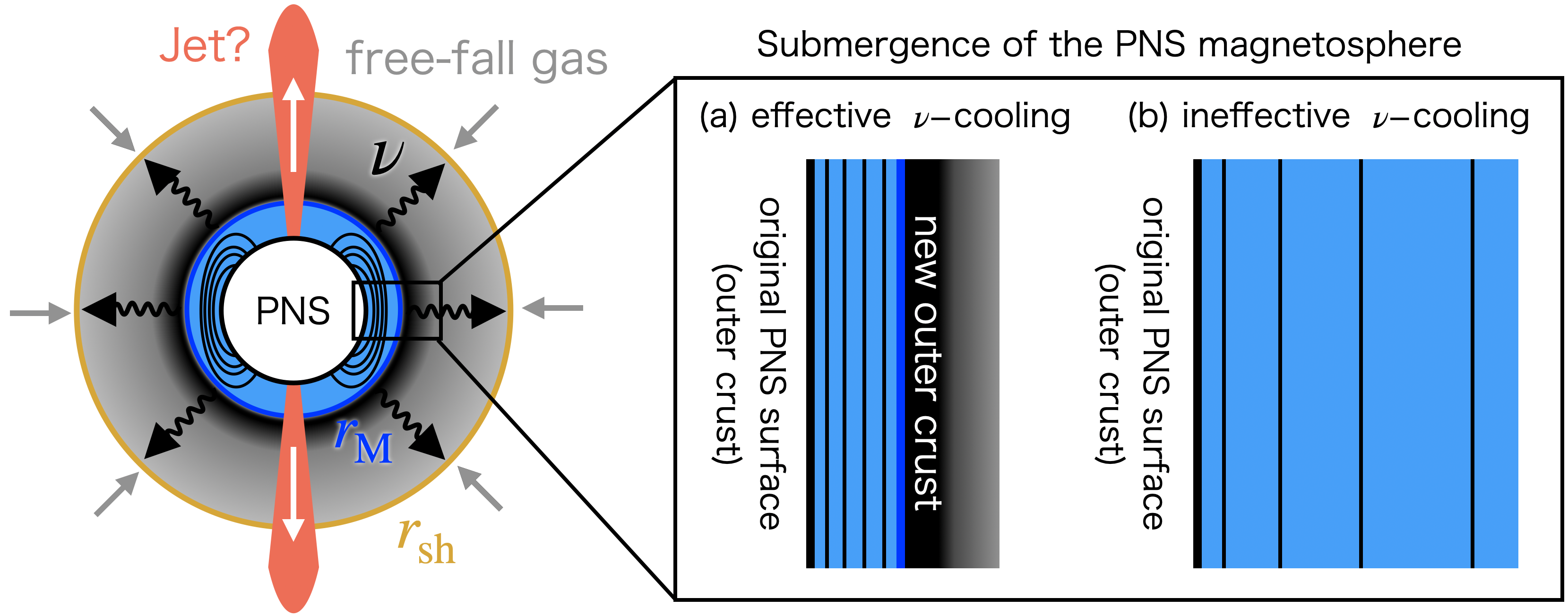}
\caption{A schematic view of the supernova fallback.
The white circle in the left figure denotes the PNS with a radius of $r_{\rm PNS}$.
The gray arrows stand for the velocity vector of the free-falling gas, and black wavy lines with arrows indicate the neutrinos propagating outward.
We denote the magnetospheric and shock radii as $r_{\rm M}$ and $r_{\rm sh}$, respectively, in terms of the spherical radius $r$. 
The overall structure of the supernova fallback can be divided into free-fall region ($r>r_{\rm sh}$), post-shock region ($r_{\rm M}<r<r_{\rm sh}$), and the PNS magnetosphere ($r_{\rm PNS}<r<r_{\rm M}$).
The shadow in the post-shock region represents the gas density and temperature.
The right two panels are enlarged views around $r=r_{\rm M}$.
A new crust forms above the PNS magnetosphere when neutrino cooling is efficient (panel a), whereas no crust forms when neutrino cooling is inefficient (panel b).
\label{fig:figure1}}
\end{figure*}

As illustrated in the left panel of Figure \ref{fig:figure1}, we consider that the fallback accretion onto a magnetized PNS consists of three regions \citep{Torres2016}: the free-fall region ($r>r_{\rm sh}$), the post-shock region ($r_{\rm M}<r<r_{\rm sh}$), and the PNS magnetosphere ($r_{\rm PNS}<r<r_{\rm M}$).
Here, we denote the magnetospheric and shock radii by $r_{\rm M}$ and $r_{\rm sh}$, respectively, in terms of the spherical radius $r$.
We write the PNS radius as $r_{\rm PNS}$.
For $r>r_{\rm sh}$, the gas accretes toward the PNS at a free-fall velocity.
For $r_{\rm M}<r<r_{\rm sh}$, a strong shock forms and the fallback material slowly settles to the surface of $r=r_{\rm M}$.
For $r_{\rm PNS}<r<r_{\rm M}$, the PNS's magnetic field is confined by the post-shock flows.
When $\dot{M}_{\rm fb}$ is very high, neutrino emission occurs at $r\sim r_{\rm M}$.
Such neutrino emission releases the gravitational energy liberated at $r\sim r_{\rm M}$.
The two panels on the right are enlarged views around $r=r_{\rm M}$.
These panels describe the cases where neutrino cooling is effective (panel a) and ineffective (panel b).
If the neutrino cooling is effective enough for the new crust to form, then the PNS magnetosphere is submerged.
Additionally, if the PNS rotation is significant, a bipolar jet powered by the magnetic field \citep[][]{Parfrey2017} or neutrinos \citep[][]{Qian1996} will emerge around the rotation axis.

Neutrino emissions affect the dynamics of the supernova fallback.
\citet{Chevalier1989} found steady solutions in which the gravitational energy released at the PNS surface is lost through neutrino cooling in the post-shock region \citep[see also][]{Houck1991}.
The steady solution is later confirmed by one-dimensional general relativistic Boltzmann neutrino-hydrodynamic simulations by \citet{Akaho2024}.
Since these studies considered the non-magnetzed PNS, it is still not understood how the neutrino emission affects the dynamics of the supernova fallback onto a magnetized PNS.

In three dimensions, convection can influence the submergence of the PNS magnetic fields through convective diffusion.
\citet{Bernal2013} presents a set of three-dimensional magnetohydrodynamic (MHD) simulations examining the interaction between fallback matter and an isolated magnetic arcade, suggesting that the post-shock region is marginally convectively stable \citep[see also][]{Bernal2010}.
Therefore, one- and two-dimensional models that neglect the effects of convection remain reasonable.
Since this pioneering three-dimensional study explored only a limited parameter space, comprehensive surveys using one- or two-dimensional models are important for identifying the conditions under which the submergence of the magnetic field occurs.

In this study, we perform one-dimensional general relativistic MHD (GRMHD) simulations of the supernova fallback onto a magnetized PNS over wide parameter ranges.
Our simulations take neutrino cooling into account.
We study how the neutrino cooling affects the dynamics of fallback material for various PNS magnetic field strengths.
Using numerical models, we derive a necessary condition for the submergence of the PNS magnetic field.

This paper is organized as follows.
In Section \ref{sec:overview}, we briefly review the steady solution without magnetic fields in \cite{Chevalier1989}, which provides a key concept of the fallback process.
We will present the numerical methods and models in Section \ref{sec:method} and show the numerical results in Section \ref{sec:result}. 
Section \ref{sec:implications} is devoted to the discussion of the diversity of young isolated NSs.
We discuss the limitations of our models in Section \ref{sec:limitation}.
Finally, we give our conclusion in Section \ref{sec:conclusion}.
Hereafter, the speed of light $c$ and the gravitational constant $G$ are normalized to $1$ unless otherwise specified.
We represent the PNS mass as $M_{\rm PNS}$ and the gravitational radius as $r_{\rm g}=M_{\rm PNS}$.

{\section{Overview of Chevalier 1989}\label{sec:overview}}

We briefly review the steady solution of the supernova fallback onto a non-magnetized PNS \citep[][]{Chevalier1989}, which provides a guideline of our analysis.
Since neutrino cooling is effective only in the vicinity of the PNS surface, the post-shock flow will be adiabatic over most of the accretion flow.
We consider situations in which radiation pressure dominates gas pressure because of high accretion rates.
To take into account the effect, we take an effective specific heat ratio of $\Gamma=4/3$ for the equation of state (EoS).
Additionally, since local thermodynamic equilibrium will be achieved in the flows, we regard the radiation and gas temperatures as equal.
Thus, we calculate the gas temperature as $T=(3p/a_{\rm rad})^{1/4}$, where $a_{\rm rad}$ is the radiation constant.
The electron-positron pair creation affects the pressure in reality, but we ignore it for simplicity.
We will discuss the effect in Section \ref{sec:limitation}.

From the Rankine-Hugoniot relations, the gas density $\rho$, pressure $p$, and velocity $v$ in the post-shock region for a standing shock at the radius of $r_{\rm sh}$ are described as functions of radius $r$:
\begin{eqnarray}
  \rho(r)&=&
  7\rho_{\rm pre}\left(\frac{r}{r_{\rm sh}}\right)^{-3},
  \label{eq:shock_rho}\\
  p(r)&=&
  \frac{6}{7}\rho_{\rm pre}v_{\rm ff}^2(r_{\rm sh})\left(\frac{r}{r_{\rm sh}}\right)^{-4},
   \label{eq:shock_pres}\\
  v(r)&=&
  \frac{v_{\rm ff}(r_{\rm sh})}{7}\left(\frac{r}{r_{\rm sh}}\right),
   \label{eq:shock_vel}
\end{eqnarray}
where $\rho_{\rm pre}=\dot{M}_{\rm fb}/(4\pi r_{\rm sh}^2v_{\rm ff}(r_{\rm sh}))$ is the pre-shock gas density at $r=r_{\rm sh}$, and $v_{\rm ff}$ is the free-fall velocity from infinity.

The steady-state solution is obtained by assuming that all gravitational energy released by accretion is lost through the neutrino emission:
\begin{eqnarray}
    \frac{M_{\rm PNS}\dot{M}_{\rm in}}{r_{\rm PNS}}=4\pi r_{\rm PNS}^2H\dot{q},
    \label{eq:energy_eq_anal}
\end{eqnarray}
where $\dot{q}$ is the cooling rate due to the neutrino emission.
\cite{Chevalier1989} adopted $H=r_{\rm PNS}^2p/({M_{\rm PNS}\rho})=(12/49)r_{\rm PNS}$ for the fiducial width of the layer in which neutrino cooling works effectively.
From Equations (\ref{eq:shock_rho}) - (\ref{eq:energy_eq_anal}), we get $r_{\rm sh}$ as a function of $\dot{M}_{\rm fb}$.

\begin{figure*}[tb]
\centering
\includegraphics[width=\linewidth]{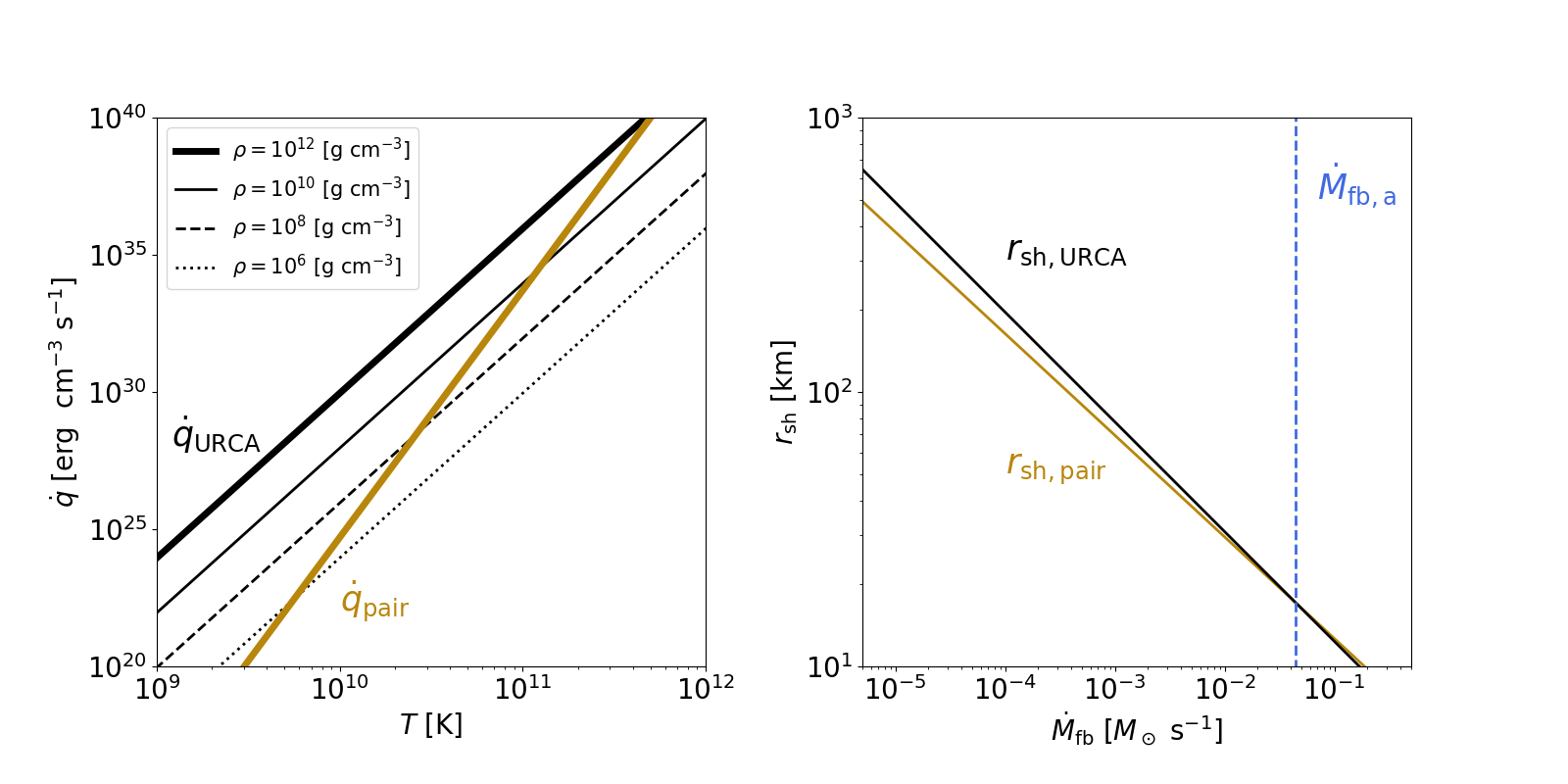}
\caption{
Left panel: cooling rate due to the neutrino emission against the gas temperature. 
We employ $\rho=10^{6},~10^{8},~10^{10}$, and $10^{12}~{\rm cm^{3}~s^{-1}}$.
Right panel: steady solutions of the shock radius as a function of fallback accretion rate.
Here, we adopt $r_{\rm PNS}=10~{\rm km}$ and $M_{\rm PNS}=1.4M_\odot$ (see Equations \ref{eq:pair} and \ref{eq:URCA}).
The vertical dashed line denotes the mass accretion rate when $r_{\rm sh,pair}=r_{\rm sh,URCA}$ ($\dot{M}_{\rm fb,a}$).
\label{fig:figure2}}
\end{figure*}

The left panel of Figure \ref{fig:figure2} shows the cooling rate as a function of the gas temperature.
We consider two processes relevant to the fallback accretion \citep{Itoh1989,Qian1996}:
\begin{eqnarray}
    \dot{q}_{\rm pair}
    &=&
    5\times10^{33}[{\rm erg~cm^{-3}~s^{-1}}]
    \left(\frac{T}{10^{11}~{\rm K}}\right)^9.
    \label{eq:pair_neu}\\
    \dot{q}_{\rm URCA}
    &=&
    9\times10^{29}[{\rm erg~cm^{-3}~s^{-1}}]
    \left(\frac{\rho}{10^6~{\rm g~cm^{-3}}}\right)
    \left(\frac{T}{10^{11}~{\rm K}}\right)^6,
    \label{eq:URCA_neu}
\end{eqnarray}
The pair neutrino process is $e^-+e^+\rightarrow \nu_{\rm e}+\bar{\nu}_{\rm e}$, and the URCA processes are $p+e^-\rightarrow n+\nu_{\rm e}$ and $n+e^+\rightarrow p+\bar{\nu}_{\rm e}$.
For $\rho =10^{8}~{\rm g~cm^{-3}}$, $\dot{q}_{\rm pair}$ is dominant over $\dot{q}_{\rm URCA}$ for $T\gtrsim 3\times10^{10}~{\rm K}$.

The right panel of Figure \ref{fig:figure2} displays the analytical solutions of the shock radius $r_{\rm sh,pair}$ (gold line) and $r_{\rm sh,URCA}$ (black line) against $\dot{M}_{\rm fb}$:
\begin{eqnarray}
    r_{\rm sh,pair}&=&3.4\times10^{2}~[{\rm km}]
    \left(\frac{r_{\rm PNS}}{10~{\rm km}}\right)^{40/27}
    \left(\frac{M_{\rm PNS}}{1.4~{\rm M}_\odot}\right)^{-1/27}
    \left(\frac{\dot{M}_{\rm fb}}{10^{-5}~{\rm M_\odot~s^{-1}}}\right)^{-10/27},
    \label{eq:pair}\\
    r_{\rm sh,URCA}&=&4.3\times10^{2}~[{\rm km}]
    \left(\frac{r_{\rm PNS}}{10~{\rm km}}\right)^{4/3}
    \left(\frac{M_{\rm PNS}}{1.4~{\rm M}_\odot}\right)^{1/5}
    \left(\frac{\dot{M}_{\rm fb}}{10^{-5}~{\rm M_\odot~s^{-1}}}\right)^{-2/5},
    \label{eq:URCA}
\end{eqnarray}
for $\dot{q}=\dot{q}_{\rm pair}$ and $\dot{q}=\dot{q}_{\rm URCA}$, respectively.
Because both $\dot{q}_{\rm pair}$ and $\dot{q}_{\rm URCA}$ increase with $\dot{M}_{\rm fb}$, both $r_{\rm sh,pair}$ and $r_{\rm sh,URCA}$ decrease with $\dot{M}_{\rm fb}$.
The vertical dashed line indicates the accretion rate for $r_{\rm sh,pair}=r_{\rm sh,URCA}$ ($\dot{M}_{\rm fb,a}\approx 4\times10^{-2}~{\rm M_\odot ~s^{-1}}$).
When $\dot{M}_{\rm fb}<\dot{M}_{\rm fb,a}$, we get the relation of $r_{\rm sh,pair}<r_{\rm sh,URCA}$.
This relation indicates $\dot{q}_{\rm pair}>\dot{q}_{\rm URCA}$ for $\dot{M}_{\rm fb}<\dot{M}_{\rm fb,a}$.
On the other hand, $\dot{q}_{\rm pair}<\dot{q}_{\rm URCA}$ holds for $\dot{M}_{\rm fb}>\dot{M}_{\rm fb,a}$.
The shock radii are smaller than the PNS radius for $\dot{M}_{\rm fb}\gtrsim 10^{-1}~{\rm M_\odot~s^{-1}}$, indicating that the free-fall region reaches the PNS surface.
In the present study, we investigate the supernova fallback with a mass accretion rate of $\dot{M}_{\rm fb}=10^{-(5-2)}~{\rm M_\odot~s^{-1}}$ (see Section \ref{sec:numrical_model} for details).

{\section{Numerical Method} \label{sec:method}}

We numerically solve the GRMHD equations while considering neutrino cooling in Schwarzschild polar coordinates $(t,r,\theta,\phi)$.
These simulations are one-dimensional in spherical symmetry, with the magnetic field depending only on the radius \citep[][]{Cumming2001}.
We use the numerical code {\tt UWABAMI} \citep[e.g.,][]{Takahashi2017}.
We represent space-time and space components as Greek and Latin suffixes, respectively.

{\subsection{Basic equations}\label{sec:basic_equations}}

The equations for the time-evolution of the GRMHD are as follows:
\begin{eqnarray}
    \nabla_\mu\left(\rho u^\mu\right)&=&0,
    \label{eq:mass_cons}
    \\
    \nabla_\mu\left(T^{\mu\nu}\right)&=&\sqrt{-g}Q^\nu,
    \label{eq:momentum_cons_gas}
    \\
    \partial_t\left(\sqrt{-g}B^i\right)&=&-\partial_j\left\{\sqrt{-g}\left(b^i u^j-b^j u^i\right)\right\},
    \label{eq:induction_eq}
\end{eqnarray}
where $u^\mu$ is the four-velocity of the gas, $B^i$ is the magnetic field vector in the laboratory frame, $b^\mu$ is the magnetic four-vector in the fluid frame, and $g={\rm det}(g_{\mu\nu})$ is the determinant of the metric.
The energy-momentum tensor of the ideal MHD is given by
\begin{eqnarray}
    {T}^{\mu\nu}&=&
    \left(\rho+e+p+\frac{b^2}{8\pi}+p_{\rm mag}\right)u^{\mu}u^{\nu}
    +\left(p+p_{\rm mag}\right)g^{\mu\nu}
    -\frac{b^\mu b^\nu}{4\pi},
    \label{eq:Tmunu_gas}
\end{eqnarray}
where $e=p/(\Gamma-1)$, $b^2=b^\mu b_\mu$, and $p_{\rm mag}=b^2/8\pi$ is the magnetic pressure in the fluid frame.
We express the source term accounting for the neutrino emission as
\begin{eqnarray}
    Q^\mu&=&-\left(\dot{q}_{\rm pair}+\dot{q}_{\rm URCA}\right)u^\mu.
    \label{eq:optically_thin_cooling}
\end{eqnarray}

The numerical methods used to solve the above equations are as follows.
We utilize a harmonic mean proposed by \citet{vanLeer1977} for the reconstruction and adopt the Harten–Lax–van Leer \citep[HLL; ][]{Harten1983} scheme to evaluate the numerical flux.
We perform the time integration of the advection term using an explicit second-order Runge-Kutta scheme.
On the other hand, we implicitly solve the source term responsible for the neutrino cooling $Q^\mu$ with the Newton-Raphson method.
Since the gas velocity in the region where neutrino cooling is pronounced is much smaller than the speed of light, we take $Q^i=0$.
This approach enables us to solve the source term using the Newton-Raphson method for a single variable, $p$, alone, rather than for four variables, $p$ and $u^i$, thereby reducing computational cost.
Although we do not solve the energy and momentum transport due to neutrino scattering and absorption, these processes do not affect our conclusions within the density range treated in the present study (see Appendix \ref{sec:sca_abs}).

\begin{deluxetable}{cccccc}[t]
\tablecaption{Parameters for numerical models\label{tab:table1}}
\tablehead{
\colhead{Parameters}
&\colhead{$B_{\rm PNS}$}
&\colhead{$\rho_{\rm out}$}
&\colhead{$\dot{M}_{\rm fb}$}
&\colhead{$t_{\rm shock}$}\\
\colhead{Unit}
&\colhead{$[\rm G]$}
&\colhead{$[\rm g~cm^{-3}]$}
&\colhead{$[\rm M_\odot~s^{-1}]$}
&\colhead{$[\rm s]$}
}
\startdata
{\tt BnMm5R10}    &  $0$      & $10^2$ & $10^{-5}$ & $3.48\times10^{-2}$\\
{\tt BnMm4R10}    &  $0$      & $10^3$ & $10^{-4}$ & $3.49\times10^{-2}$\\
{\tt BnMm3R10}    &  $0$      & $10^4$ & $10^{-3}$ & $3.48\times10^{-2}$\\
{\tt BnMm2R10}    &  $0$      & $10^5$ & $10^{-2}$ & $3.48\times10^{-2}$\\
{\tt B13Mm5R10}   &  $10^{13}$& $10^2$ & $10^{-5}$ & $3.49\times10^{-2}$\\
{\tt B13Mm4R10}   &  $10^{13}$& $10^3$ & $10^{-4}$ & $3.49\times10^{-2}$\\
{\tt B13Mm3R10}   &  $10^{13}$& $10^4$ & $10^{-3}$ & $3.49\times10^{-2}$\\
{\tt B13Mm2R10}   &  $10^{13}$& $10^5$ & $10^{-2}$ & $3.48\times10^{-2}$\\
{\tt B14Mm5R10}   &  $10^{14}$& $10^2$ & $10^{-5}$ & $3.54\times10^{-2}$\\
{\tt B14Mm4R10}   &  $10^{14}$& $10^3$ & $10^{-4}$ & $3.51\times10^{-2}$\\
{\tt B14Mm3R10}   &  $10^{14}$& $10^4$ & $10^{-3}$ & $3.50\times10^{-2}$\\
{\tt B14Mm2R10}   &  $10^{14}$& $10^5$ & $10^{-2}$ & $3.49\times10^{-2}$\\
{\tt B15Mm4R10}   &  $10^{15}$& $10^3$ & $10^{-4}$ & $3.61\times10^{-2}$\\
{\tt B15Mm3R10}   &  $10^{15}$& $10^4$ & $10^{-3}$ & $3.54\times10^{-2}$\\
{\tt B15Mm2R10}   &  $10^{15}$& $10^5$ & $10^{-2}$ & $3.51\times10^{-2}$\\
{\tt BnMm4R20}    &  $0$      & $10^3$ & $10^{-4}$ & $3.49\times10^{-2}$\\
{\tt BnMm3R20}    &  $0$      & $10^4$ & $10^{-3}$ & $3.49\times10^{-2}$\\
{\tt BnMm2R20}    &  $0$      & $10^5$ & $10^{-2}$ & $3.48\times10^{-2}$\\
{\tt B13Mm4R20}   &  $10^{13}$& $10^3$ & $10^{-4}$ & $3.50\times10^{-2}$\\
{\tt B13Mm3R20}   &  $10^{13}$& $10^4$ & $10^{-3}$ & $3.50\times10^{-2}$\\
{\tt B13Mm2R20}   &  $10^{13}$& $10^5$ & $10^{-2}$ & $3.48\times10^{-2}$\\
{\tt B14Mm4R20}   &  $10^{14}$& $10^3$ & $10^{-4}$ & $3.59\times10^{-2}$\\
{\tt B14Mm3R20}   &  $10^{14}$& $10^4$ & $10^{-3}$ & $3.53\times10^{-2}$\\
{\tt B14Mm2R20}   &  $10^{14}$& $10^5$ & $10^{-2}$ & $3.50\times10^{-2}$\\
{\tt B15Mm3R20}   &  $10^{15}$& $10^4$ & $10^{-3}$ & $3.72\times10^{-2}$\\
{\tt B15Mm2R20}   &  $10^{15}$& $10^5$ & $10^{-2}$ & $3.59\times10^{-2}$
\enddata
\tablecomments{
The model names are shown in the first column: ``{\tt BXX}'' means the magnetic field strength at the PNS surface of $10^{{\rm XX}}~{\rm G}$ (``{\tt Bn}'' is the case of $B_{\rm PNS}=0$), ``{\tt MmX}'' denotes the fallback mass accretion rate of $10^{-{\rm X}}~{\rm M_\odot~s^{-1}}$, and ``{\tt RXX}'' represents the PNS radius of ${\rm XX}~{\rm km}$.
We present the magnetic field strength at the PNS surface ($B_{\rm PNS}$), the gas density at $r=r_{\rm out}$ ($\rho_{\rm out}$), the mass accretion rate at $r=r_{\rm out}$ ($\dot{M}_{\rm fb}=-4\pi r_{\rm out}^2\rho_{\rm out}u^r$), and the time when the accretion shock forms ($t_{\rm shock}$).}
\end{deluxetable}

{\subsection{Numerical setup}\label{sec:numrical_model}}

We set the radius of the outer boundary to $r_{\rm out}=10^3~{\rm km}$.
The computational domain consists of $[r_{\rm PNS},r_{\rm out}]$.
The number of numerical grid points is $N_r=16384$.
The radius of the grid exponentially increases with $e^{x_a}$, where $x_a=\ln r_{\rm PNS}+a\times dX$, $a$ is the radial grid index, and $dX=(\ln r_{\rm out}-\ln r_{\rm PNS})/N_r$.
The minimum grid width is $\Delta r\approx 280~{\rm cm}$, which is much smaller than the pressure scale height near the PNS surface $\sim 10^5 {\rm cm}$ for $\rho\sim 10^{10}~{\rm g~cm^{3}}$ and $T\sim 10^{11}~{\rm K}$.

Table \ref{tab:table1} presents the magnetic field strength at the PNS surface ($B_{\rm PNS}$), the gas density at $r=r_{\rm out}$ ($\rho_{\rm out}$), the mass accretion rate at $r=r_{\rm out}$ ($\dot{M}_{\rm fb}=-4\pi r_{\rm out}^2 \rho u^r$), and the time at which the accretion shock forms ($t_{\rm shock}$).
We will provide an explanation of $t_{\rm shock}$ in Section \ref{sec:time_evol}.

We adopt typical values for the PNS mass and radius.
The PNS contracts and cools down by neutrino emission after its birth.
This contraction phase is considered to last for $\sim 1-10~{\rm s}$.
The mass and radius of the PNS typically lie in $\approx 1-2~M_\odot$ and $\approx 10$–$20~{\rm km}$, respectively \citep[e.g.,][]{Fischer2010,Nagakura2022}.  
We take $M_{\rm PNS}=1.4~M_\odot$ as a fiducial value.
We adopt $r_{\rm PNS}=10$ and $20~{\rm km}$ and keep it fixed in the simulations for simplicity.
We first show the results for $r_{\rm PNS} = 10~{\rm km}$.
We present the results for $r_{\rm PNS} = 20~{\rm km}$ in Section \ref{sec:dependence_on_rPNS} and explain how the results depend on $r_{\rm PNS}$.

We consider a magnetized and non-rotating PNS.
The PNS initially has a magnetic field given by $B^{(\theta)}=B_{\rm init}=B_{\rm PNS}(r/r_{\rm PNS})^{-3}$ and $B^r=B^\phi=0$.
The parentheses in the indices of physical quantities denote the values in the static observer frame.
This configuration models the fallback material accreting perpendicularly to the dipole magnetic field.
We ignore the effects of the large-scale field geometry and magnetic tension.
We will discuss these effects in Section \ref{sec:limitation}.
We will also discuss how the PNS rotation affects the submergence process in Section \ref{sec:implications}.

We initially set the atmosphere in an MHD equilibrium to maintain the radial profile of $B^{(\theta)}$.
From the relation of $dp_{\rm mag}/dr=-M_{\rm PNS}\rho/r^2$, we can describe the initial gas density as $\rho=(3/4\pi)B_{\rm PNS}^2(r_{\rm PNS}/r_{\rm g})(r/r_{\rm PNS})^{-5}$.
We set the initial pressure ratio to $p/p_{\rm mag}=0.01$.
We also investigate the case of $B_{\rm PNS}=0$ for comparison.
In this case, we take $\rho=10^{-4}\rho_{\rm out}(r/r_{\rm PNS})^{-1/(\Gamma-1)}$ and $p=10^{-4}\rho_{\rm out}[(\Gamma-1)/\Gamma](r_{\rm g}/r_{\rm PNS})(r/r_{\rm PNS})^{-\Gamma/(\Gamma-1)}$, which satisfy $dp/dr=-M_{\rm PNS}\rho/r^2$.
In all models, we initially set $u^i=0$.

In the ghost cells inside $r_{\rm PNS}$, we fix $B^i$ to the initial value and set $\rho$ and $p$ according to the zero-gradient boundary condition.
\footnote{
The zero-gradient boundary condition at the inner boundary results in a discontinuity with the hydrostatic gas.
To investigate the effect of the discontinuity, we conducted a non-accreting test.
We confirm that the discontinuity does not affect the dynamics of the supernova fallback.
Therefore, the inner boundary condition adopted in this study will not affect our conclusion.}
We adopt the mirror boundary condition for $u^r$.
The electrical conductivity inside the PNS is sufficiently high that the electric field can be considered negligible \citep{Igoshev2021}.
Therefore, we set the numerical flux in the radial component of Equation (\ref{eq:induction_eq}) to 0 at $r=r_{\rm PNS}$.
From the outer boundary at $r=r_{\rm out}$, we inject non-magnetized gas with a density of $\rho_{\rm out}$, a velocity of $v_{\rm ff}(r_{\rm out})$, and a temperature of $5\times10^7~{\rm K}$ as the fallback material.

We adopt numerical floors for density and pressure to improve numerical stability: $\rho_{\rm fl}=10^{-4}\rho_{\rm out}(r/r_{\rm PNS})^{-3}$ and $p_{\rm fl}=10^{-6}\rho_{\rm out}(\Gamma-1)(r/r_{\rm PNS})^{-4}$, respectively.
In regions where the kinetic or magnetic energy is much greater than the thermal energy, small errors can lead to a negative pressure. We can avoid the problem with these floors.

{\section{Numerical Results} \label{sec:result}}

\subsection{Time evolution of general structures \label{sec:time_evol}}

\begin{figure*}[tb]
\centering
\includegraphics[width=\linewidth]{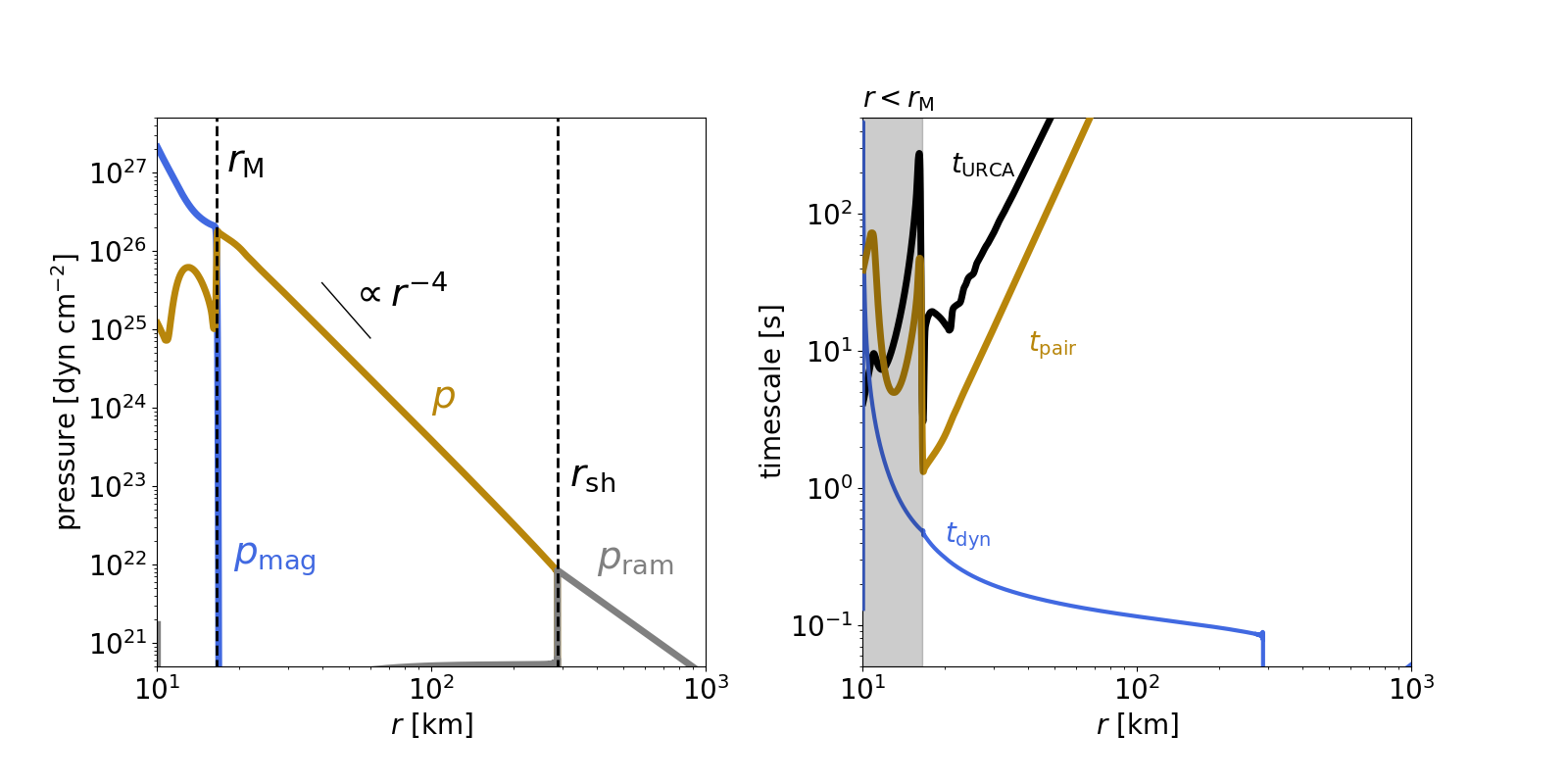}
\caption{
Left: radial profiles of three different pressures: the magnetic pressure $p_{\rm mag}$ (blue), the radiation pressure $p$, and the ram pressure $p_{\rm ram}$.
Dashed lines indicate the magnetospheric radius $r_{\rm M}$ and the shock radius $r_{\rm sh}$.
The thin line represents the radial dependence of $p$ in Equation ({\ref{eq:shock_pres}}).
Right: various time scales against radius.
We express the dynamical time scale as $t_{\rm dyn}=r/v^{(r)}$ and the cooling time scales due to the pair process as $t_{\rm pair}=e/\dot{q}_{\rm pair}$ and the URCA process as $t_{\rm URCA}=e/\dot{q}_{\rm URCA}$.
The shaded region stands for a region of $r<r_{\rm M}$.
Both panels are results in the fiducial model, {\tt B14Mm5R10}, at $t=0.118~{\rm s}$.
\label{fig:figure3}}
\end{figure*}

We first present the accretion structure in our fiducial model, {\tt B14Mm5R10}.
Subsequently, we describe the remaining models of $r_{\rm PNS}=10~{\rm km}$ in comparison with the fiducial model.
We show the results of $r_{\rm PNS} = 20~{\rm km}$ in Section \ref{sec:dependence_on_rPNS}.

The left panel of Figure \ref{fig:figure3} presents the radial profiles of three different pressures in the fiducial model at $t=0.118~{\rm s}$: $p_{\rm mag}$ (blue), $p$ (gold), and $p_{\rm ram}=\rho(v^{(r)})^2$ (gray), where $v^{(r)}=u^{(r)}/u^{(t)}$.
As mentioned in Figure \ref{fig:figure1}, there are three distinct regions: the free-fall region, the post-shock region, and the PNS magnetosphere.
These regions correspond to where $p_{\rm ram}$, $p$, and $p_{\rm mag}$ are the highest among the three, respectively.
We define the magnetospheric radius $r_{\rm M}$ and the shock radius $r_{\rm sh}$ as the radii at which $p_{\rm mag}=p$ and $p=p_{\rm ram}$, respectively.
We then obtain $r_{\rm M}=20~{\rm km}$ and $r_{\rm sh}=300~{\rm km}$ at this time (see dashed lines).

The PNS magnetic field decelerates the accreting gas, driving an accretion shock.
We define the time at which the accretion shock forms as $t_{\rm shock}$.
More specifically, we define $t_{\rm shock}$ as the time when a region satisfying $p>\max{(p_{\rm mag},p_{\rm ram})}$ and $p_{\rm rad}>10^{24}~{\rm dyn~cm^{-2}}$ first appears.
As presented in Table \ref{tab:table1}, $t_{\rm shock}$ is almost the same for all models, indicating that $t_{\rm shock}$ is insensitive to the initial field strength of the PNS.

The thin solid line in the left panel of Figure \ref{fig:figure3} represents the radial dependence given in Equation (\ref{eq:shock_pres}).
The resulting $\rho$, $p$, and $v^{(r)}$ agree with Equations (\ref{eq:shock_rho}) - (\ref{eq:shock_vel}).
By comparing the thin line with the gold line, we find that the resulting pressure $p$ follows the dependence of $r^{-4}$.
At $r=r_{\rm sh}$, the jump condition of $p=(6/7)p_{\rm ram}$ is satisfied.
Although not shown in this panel, $\rho$ and $v^{(r)}$ are also consistent with analytical solutions (Equations \ref{eq:shock_rho} and \ref{eq:shock_vel}).
These analytic solutions are valid for a standing shock.
Therefore, when the shock expansion speed $v_{\rm sh} = dr_{\rm sh}/dt$ is not negligible compared to $|v^{(r)}|$ at $r = r_{\rm sh}$, the resulting $\rho$, $p$, and $v^{(r)}$ deviate from the analytic solutions.

The right panel of Figure \ref{fig:figure3} shows the various time scales versus radius in the fiducial model at $t=0.118~{\rm s}$: the dynamical time scale $t_{\rm dyn}=r/v^{(r)}$ (blue), and cooling time scales due to the pair process $t_{\rm pair}=e/\dot{q}_{\rm pair}$ (gold) and the URCA process $t_{\rm URCA}=e/\dot{q}_{\rm URCA}$ (black).
At this time, neutrino cooling is ineffective.
The shaded region represents a region of $r<r_{\rm M}$.
We express the time scale of the neutrino cooling as $t_{\rm cool}=\min(t_{\rm pair},t_{\rm URCA})$.
For $r>r_{\rm M}$, the dynamical time scale is shorter than the cooling time scale, indicating the ineffective neutrino cooling.
This panel also shows $t_{\rm pair}<t_{\rm URCA}$, which implies $\dot{q}_{\rm pair}>\dot{q}_{\rm URCA}$.

\begin{figure*}[tb]
\centering
\includegraphics[width=\linewidth]{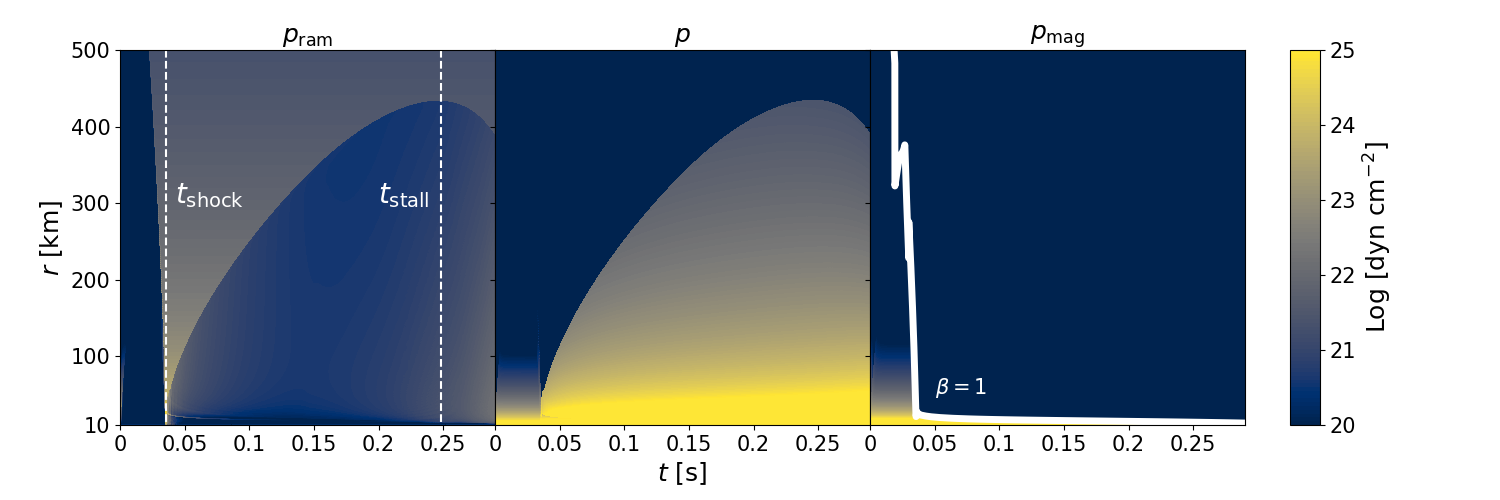}
\caption{
The time-sequenced images of $p_{\rm ram}$ (left), $p$ (middle), and $p_{\rm mag}$ (right) in the fiducial model, {\tt B14Mm5R10}.
Dashed lines in the left panel represent the shock formation time $t_{\rm shock}$ and the shock stalling time $t_{\rm stall}$.
For $t_{\rm shock}<t<t_{\rm stall}$, the shock expands over time.
However, it stalls at $t=t_{\rm stall}$ and recedes for $t>t_{\rm stall}$ due to the neutrino cooling.
The white line in the right panel indicates the radius of $\beta=1$.
The region on the lower and upper sides of this line correspond to $\beta<1$ and $\beta>1$, respectively.
\label{fig:figure4}}
\end{figure*}

Figure \ref{fig:figure4} displays time-sequenced images of the radial profiles of $p_{\rm ram}$ (left), $p$ (middle), and $p_{\rm mag}$ (right) in the fiducial model.
For $t>t_{\rm shock}$, both $p_{\rm ram}$ and $p$ change discontinuously at $r=r_{\rm sh}$.
The dashed lines in the left panel indicate the times $t = t_{\rm shock}$ and $t = t_{\rm stall}$, where $t_{\rm stall}$ is the time when the expansion of $r_{\rm sh}$ stops (i.e., $v_{\rm sh} = 0$).
In the early stage ($t_{\rm shock}<t<t_{\rm stall}$), $r_{\rm sh}$ expands over time.

In the later stage ($t\ge t_{\rm stall}$), the accretion shock stalls and recedes due to the neutrino cooling.
As the accreted mass accumulates above the PNS magnetosphere, the pressure just above it increases.
According to the relation of $T=(3p/a_{\rm rad})^{1/4}$, $T$ at $r\sim r_{\rm M}$ also increases.
Since neutrino cooling is more effective at a higher temperature, $t_{\rm cool}$ can become smaller than $t_{\rm dyn}$ at $r\sim r_{\rm M}$.
The effective neutrino cooling reduces $p$ at $r\sim r_{\rm M}$, and the radiation pressure is no longer able to support gravity.
For this reason, the post-shock region collapses.
We will provide the dependence of the shock stalling time scale on $\dot{M}_{\rm fb}$ and $B_{\rm PNS}$ in Section \ref{sec:tstall}.

The line in the right panel of Figure \ref{fig:figure4} shows the radius of $\beta=1$, where $\beta=(p_{\rm ram}+p)/p_{\rm mag}$.
This radius decreases rapidly in the early time, indicating the compression of the PNS fields.
The $\beta=1$ line lies in the vicinity of the PNS surface for $t> 0.05~{\rm s}$, showing the formation of a thin magnetosphere.

We note that the neutrino cooling works effectively not around $r\sim r_{\rm PNS}$, but around $r\sim r_{\rm M}$.
We define the pressure scale height at $r=r_{\rm M}$ as $H_{\rm M}=r_{\rm M}^2p/({M\rho})$.
We find that $t_{\rm cool}$ is shorter than $t_{\rm dyn}$ for $r_{\rm M}\lesssim r\lesssim r_{\rm M}+H_{\rm M}$ when $t=t_{\rm stall}$.
Additionally, the gas density at $r\sim r_{\rm M}$ rapidly increases for $t>t_{\rm stall}$ due to the neutrino cooling, potentially leading to the formation of a new crust that confines the PNS magnetosphere.
The increase in $\rho$ for $t>t_{\rm stall}$ implies that the shock stalling time scale corresponds to the waiting time for the new crust formation.
We will discuss this point in more detail in Section \ref{sec:tstall}.

In models with a magnetized PNS and low mass-accretion rates ($B_{\rm PNS} > 0$ and $\dot{M}_{\rm fb} = 10^{-(5-4)}~{\rm M}_\odot~{\rm s}^{-1}$), the results are very similar to those of the fiducial model.
However, in models with high mass-accretion rates ($\dot{M}_{\rm fb} = 10^{-(3-2)}~{\rm M}_\odot~{\rm s}^{-1}$) or in non-magnetized PNS models ($B_{\rm PNS} = 0$), the results differ from those of the fiducial model.

In models with high mass accretion rates, neutrino cooling is effective even at $t\sim t_{\rm shock}$.
A higher $\dot{M}_{\rm fb}$ leads to larger $\rho$ and $T$ in the post-shock region, resulting in a greater $\dot{q}$.
As a result, $t_{\rm cool}$ can be shorter than $t_{\rm dyn}$.
We also find $t_{\rm URCA}\sim t_{\rm pair}$ in these models, which implies that both the URCA and pair processes are equally important.
Such effective neutrino cooling leads to a reduction in the effective adiabatic index of the gas.
Therefore, the resulting $\rho$, $p$, and $v^{(r)}$ are slightly steeper than $\propto r^{-3}$, $\propto r^{-4}$, and $\propto r$, respectively.

In the non-magnetized cases, the post-shock region is formed in $r_{\rm PNS}<r<r_{\rm sh}$, while the free-fall region is located in $r>r_{\rm sh}$.
The cooling time scale is shorter than the dynamical time scale for $r_{\rm PNS}\lesssim r\lesssim r_{\rm PNS}+H$ at $t=t_{\rm stall}$, which is consistent with the assumption in \citet{Chevalier1989} (see Section \ref{sec:overview}).

We had to stop our calculation around $t \approx 1.2 t_{\rm stall}$ because the pressure scale height $H_{\rm M}$ or $H$ became smaller than $\Delta r$ due to a significant increase in $\rho$.
This increase in $\rho$ makes it impossible to resolve the stratified atmosphere.

\subsection{Magnetospheric and shock radii \label{sec:rM_rsh}}

\begin{figure*}[tb]
\centering
\includegraphics[width=0.5\linewidth]{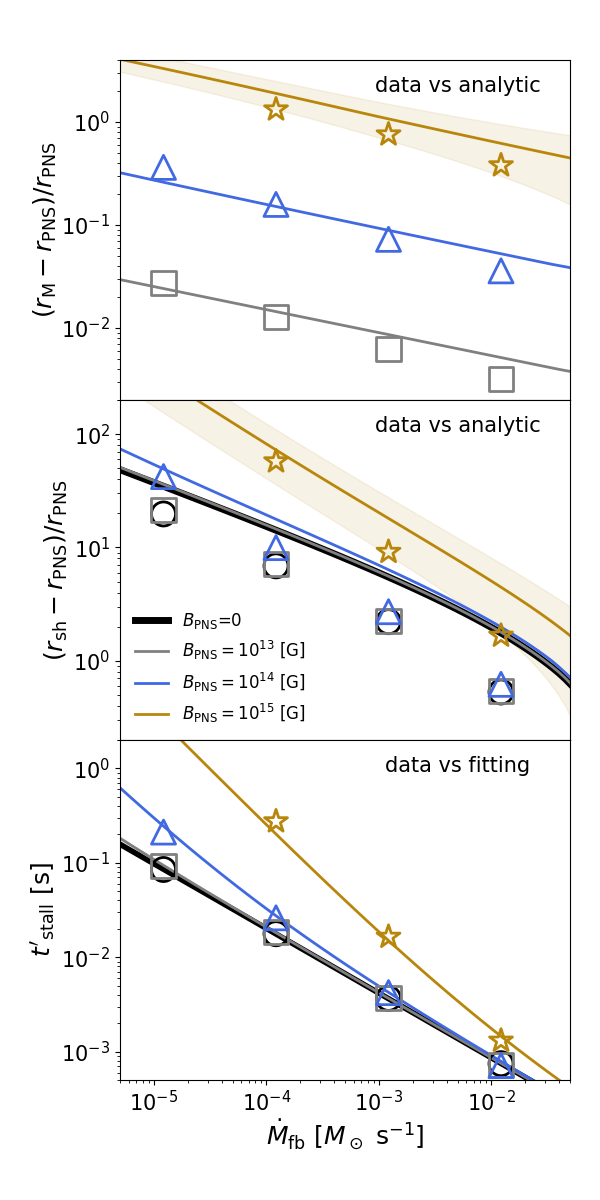}
\caption{
Top and middle: plots for the magnetospheric (top) and shock radii (middle) as a function of the fallback accretion rate.
All plots are the results when $t=t_{\rm stall}$.
The solid lines indicate the semi-analytical solutions calculated from Equations (\ref{eq:semi_ana1}), (\ref{eq:pair2}), and (\ref{eq:URCA2}).
The shaded regions in the top and middle panels denote where the deviation from the semi-analytic solution for $B_{\rm PNS}=10^{15}~{\rm G}$ is within 20\% and 50\%, respectively.
We also plot the analytical solution of $r_{\rm sh}$ for $B_{\rm PNS}=0$ (Equation \ref{eq:pair}).
Bottom: shock stalling time scale ($t'_{\rm stall}=t_{\rm stall}-t_{\rm shock}$) against the fallback accretion rate.
Solid lines represent the fitting results.
All panels are the results of $r_{\rm PNS}=10~{\rm km}$.
\label{fig:figure5}}
\end{figure*}

We show the dependence of $r_{\rm M}$ and $r_{\rm sh}$ at $t=t_{\rm stall}$ on $\dot{M}_{\rm fb}$ and $B_{\rm PNS}$.
Additionally, based on semi-analytical models, we demonstrate the following two points: (1) the magnetospheric radius is determined by the pressure balance between $p$ and $p_{\rm mag}$; (2) the shock radius is obtained from the energy balance at $r=r_{\rm M}$, which describes that the released gravitational energy is lost through the neutrino cooling.

The top panel of Figure \ref{fig:figure5} shows the parameter dependence of $r_{\rm M}$ at $t=t_{\rm stall}$.
As expected from the pressure balance, the panel demonstrates that $r_{\rm M}$ is a decreasing function of $\dot{M}_{\rm fb}$ but an increasing function of $B_{\rm PNS}$.

The middle panel of Figure \ref{fig:figure5} presents the dependence of $r_{\rm sh}$ at $t=t_{\rm stall}$ on $\dot{M}_{\rm fb}$ and $B_{\rm PNS}$.
We also plot $r_{\rm sh}$ in models of $B_{\rm PNS}=0$ for comparison.
The plots show that $r_{\rm sh}$ also decreases with $\dot{M}_{\rm fb}$ and increases with $B_{\rm PNS}$, although the dependence on $B_{\rm PNS}$ is weaker than that of $r_{\rm M}$.
When $r_{\rm M}$ is smaller, $\rho$ and $T$ just above it become higher due to stronger gravitational acceleration.
Given the parameter dependence of the  neutrino cooling (see Equations \ref{eq:pair_neu} and \ref{eq:URCA_neu}), the cooling becomes more effective, resulting in a smaller $r_{\rm sh}$.
A comparison with the non-magnetized case ($B_{\rm PNS}=0$) indicates that the PNS magnetic field significantly affects $r_{\rm sh}$ only when $B_{\rm PNS}\gtrsim10^{14}~{\rm G}$, within the range of $\dot{M}_{\rm fb}$ investigated here.

We semi-analytically derive steady solutions of $r_{\rm M}$ and $r_{\rm sh}$.
For simplicity, we use the spatial average of the magnetic field strength within the magnetosphere, denoted by $\bar{B}$.
From the magnetic flux conservation law $\int^\infty_{r_{\rm PNS}}B_{\rm init}2\pi rdr=\int^{r_{\rm M}}_{r_{\rm PNS}}\bar{B}2\pi r dr$, we then obtain the magnetic pressure at $r=r_{\rm M}$:
\begin{eqnarray}
p_{\rm mag}=\frac{\bar{B}^2}{8\pi}
=\frac{\mu^2}{2\pi}
\frac{1}{\left(r_{\rm M}^2-r_{\rm PNS}^2\right)^2}
\left(\frac{1}{r_{\rm PNS}}\right)^2,
\label{eq:mag_pres}
\end{eqnarray}
where $\mu=B_{\rm PNS}r_{\rm PNS}^3$ is the PNS magnetic moment.
Substituting Equations (\ref{eq:shock_pres}) and (\ref{eq:mag_pres}) into the pressure balance equation $p=p_{\rm mag}$, we get the following equation that relates $r_{\rm M}$ and $r_{\rm sh}$:
\begin{eqnarray}
\left[\left(\frac{\mu^2}{r_{\rm PNS}}\right)^2
-\frac{3}{7}\left(\frac{2r_{\rm g}}{r_{\rm sh}}\right)^{1/2}\dot{M}_{\rm fb}r_{\rm sh}^2\right]r_{\rm M}^4
+
\frac{6}{7}\left(\frac{2r_{\rm g}}{r_{\rm sh}}\right)^{1/2}\dot{M}_{\rm fb}r_{\rm sh}^2r_{\rm PNS}^2r_{\rm M}^2
-\frac{3}{7}\left(\frac{2r_{\rm g}}{r_{\rm sh}}\right)^{1/2}\dot{M}_{\rm fb}r_{\rm sh}^2r_{\rm PNS}^4=0.
\label{eq:semi_ana1}
\end{eqnarray}
Since the neutrino cooling is effective mainly at $r \sim r_{\rm M}$, we adopt the following energy equation instead of Equation (\ref{eq:energy_eq_anal}):
\begin{eqnarray}
    \frac{M_{\rm PNS}\dot{M}_{\rm fb}}{r_{\rm M}}=4\pi r_{\rm M}^2 H_{\rm M}\dot{q}.
\label{eq:semi_ana2}
\end{eqnarray}
Substituting $\dot{q}_{\rm pair}$ and $\dot{q}_{\rm URCA}$ into Equation (\ref{eq:semi_ana2}), we obtain the following shock radii, respectively (c.f., Equations \ref{eq:pair} and \ref{eq:URCA}):
\begin{eqnarray}
    r_{\rm sh,pair}&=&9.5\times10^{2}~[{\rm km}]
    \left(\frac{r_{\rm M}}{20~{\rm km}}\right)^{40/27}
    \left(\frac{M_{\rm PNS}}{1.4~{\rm M_\odot}}\right)^{-1/27}
    \left(\frac{\dot{M}_{\rm fb}}{10^{-5}~{\rm M_\odot~s^{-1}}}\right)^{-10/27},
    \label{eq:pair2}\\
    r_{\rm sh,URCA}&=&1.1\times10^{3}~[{\rm km}]
    \left(\frac{r_{\rm M}}{20~{\rm km}}\right)^{4/3}
    \left(\frac{M_{\rm PNS}}{1.4~{\rm M_\odot}}\right)^{1/5}
    \left(\frac{\dot{M}_{\rm fb}}{10^{-5}~{\rm M_\odot~s^{-1}}}\right)^{-2/5}
    \label{eq:URCA2}.
\end{eqnarray}

For given parameters of $\dot{M}_{\rm fb}$ and $B_{\rm PNS}$, we numerically solve Equations (\ref{eq:semi_ana1}), (\ref{eq:pair2}), and (\ref{eq:URCA2}) for $r_{\rm M}$ and $r_{\rm sh}$.
We solve these equations with a double iteration scheme.
We adopt $r_{\rm M,init}=10^3~{\rm km}$ as the initial guess of $r_{\rm M}$.
The steps of the iterative scheme are as follows:
(i) Using $r_{\rm M,init}$, we compute the shock radius from $r_{\rm sh}=\min(r_{\rm sh,pair},r_{\rm sh,URCA})$.
(ii) Based on $r_{\rm sh}$, we solve Equation (\ref{eq:semi_ana1}) for $r_{\rm M}$ using the Newton-Raphson method.
We check the convergence of $\log(r_{\rm M})$.
If a relative error of $\log(r_{\rm M})$ falls below $10^{-5}$, then we proceed to the next step.
(iii) If the obtained $r_{\rm M}$ does not coincide with the given $r_{\rm M,init}$, then we set the next guess of $r_{\rm M,init}$ and return to step (i).
The next guess is estimated from $\log(r_{\rm M,init})-10^{-6}\log(r_{\rm M,init})$.
We continue the iteration procedure until the relative accuracy $|\log(r_{\rm M,init})-\log(r_{\rm M})|/|\log(r_{\rm M})|$ reaches $0.1 \%$.

The solid lines in the top and middle panels of Figure \ref{fig:figure5} stand for the semi-analytical solutions.
We also plot the solution of $r_{\rm sh}$ for $B_{\rm PNS}=0$ (Equation \ref{eq:pair}) with a black line.
The shaded regions in the top and middle panels denote where the deviations from the semi-analytic solutions for $B_{\rm PNS}=10^{15}~{\rm G}$ remain within 20 \% and 50 \%, respectively.
These two panels show that our semi-analytical solutions are consistent with the numerical results of $r_{\rm M}$ and $r_{\rm sh}$ within the deviations of $\approx$ 20\% and $\approx$ 50\%, respectively.
We get the same results for $B_{\rm PNS}=10^{13}~{\rm G}$ and $10^{14}~{\rm G}$.

\subsection{Shock stalling time scale, $t'_{\rm stall}$\label{sec:tstall}}

As described in Section \ref{sec:time_evol}, the density just above the magnetosphere begins to increase rapidly at $t=t_{\rm stall}$ due to neutrino cooling.
This abrupt increase can potentially lead to the formation of the new crust that confines the PNS magnetosphere.
As a measure of the waiting time for the new crust formation, we define the shock stalling time scale, $t'_{\rm stall}=t_{\rm stall}-t_{\rm shock}$, and investigate its parameter dependence.
Additionally, we aim to obtain a fitting formula of $t'_{\rm stall}$ to derive a necessary condition for the submergence of the PNS field.

We first focus on the non-magnetized case as a reference (see black circles in the bottom panel of Figure \ref{fig:figure5}).
We analyze the results to find the following: when the accretion rate is high ($\dot{M}_{\rm fb}=10^{-(3-2)}~{\rm M_\odot~s^{-1}}$), the shock stalling time scale is characterized by the cooling time scale, namely $t'_{\rm stall} \approx t_{\rm cool}$.
However, when the accretion rate is low ($\dot{M}_{\rm fb}=10^{-(5-4)}~{\rm M_\odot~s^{-1}}$), $t'_{\rm stall}$ is significantly longer than $t_{\rm cool}$ due to the time delay in the onset of effective neutrino cooling.
We have confirmed that, in all models, the sound-crossing time scale across the region between $r=r_{\rm M}$ and $r=r_{\rm sh}$, $t_{\rm sound}$, is shorter than $t_{\rm cool}$ for $t\le t_{\rm stall}$.
The relation $t_{\rm sound}<t_{\rm cool}$ indicates that hydrostatic equilibrium is almost achieved in the post-shock region.

The black solid line in the bottom panel of Figure \ref{fig:figure5} denotes the fitting function for the results of the non-magnetized models.
The functional form is
\begin{eqnarray}
    t'_{\rm stall,non}(\dot{M}_{\rm fb})
    =9.8\times10^{-2}~[{\rm s}]~
    \left(\frac{\dot{M}_{\rm fb}}{{10^{-5}~\rm M_\odot~s^{-1}}}\right)^{-0.69}.
    \label{eq:tstall_Bnon}
\end{eqnarray}
Since $\dot{q}$ is large for a high $\dot{M}_{\rm fb}$, $t'_{\rm stall}$ decreases with $\dot{M}_{\rm fb}$.

The bottom panel of Figure \ref{fig:figure5} indicates that $t'_{\rm stall}$ gets longer for a stronger $B_{\rm PNS}$, as the neutrino cooling becomes more ineffective (see Section \ref{sec:rM_rsh}).
To derive a fitting function for the magnetized models, we hypothesize that $t'_{\rm stall}$ for the magnetized cases is the sum of $t'_{\rm stall,non}$ and an additional term that depends on $\dot{M}_{\rm fb}$ and $B_{\rm PNS}$, that is,
\begin{eqnarray}
    t'_{\rm stall}(\dot{M}_{\rm fb}, B_{\rm PNS}) = t'_{\rm stall,non}(\dot{M}_{\rm fb}) + f(\dot{M}_{\rm fb},B_{\rm PNS}),
    \label{eq:tstall_B}
\end{eqnarray}
where $f(\dot{M}_{\rm fb},B_{\rm PNS})$ denotes the additional term.
We then derive the fitting function for $f(\dot{M}_{\rm fb},B_{\rm PNS})$ based on the magnetized models.
In models {\tt B13Mm2R10} and {\tt B14Mm2R10}, $t'_{\rm stall}$ is slightly shorter than $t'_{\rm stall,non}$ because $r_{\rm M}\approx r_{\rm PNS}$.
Therefore, we exclude $t'_{\rm stall}$ of these models from this fitting.

The result is as follows:
\begin{eqnarray}
    f(\dot{M}_{\rm fb},B_{\rm PNS})
    =2.0\times10^{-1}~[{\rm s}]~
    \left(\frac{\dot{M}_{\rm fb}}{{10^{-5}~\rm M_\odot~s^{-1}}}\right)^{-1.2}
    \left(\frac{B_{\rm PNS}}{{10^{14}~\rm G}}\right)^{1.3}.
\end{eqnarray}
The resulting fitting functions are plotted as solid lines in the bottom panel of Figure \ref{fig:figure5}.
It turns out that the accuracy of the fitting formula is better than $\approx 96\%$.

\subsection{The dependence on the PNS radius \label{sec:dependence_on_rPNS}}

The results described above are for $r_{\rm PNS} = 10~{\rm km}$.
To clarify how the dynamics of the supernova fallback depend on $r_{\rm PNS}$, we study the case for $r_{\rm PNS} = 20~{\rm km}$ (see Table \ref{tab:table1}).

The dependence of $r_{\rm M}$ and $r_{\rm sh}$ is as follows.  
As the gravitational acceleration at the PNS surface becomes smaller with increasing $r_{\rm PNS}$, both the radiation pressure and the cooling rate decrease.
Consequently, both $r_{\rm M}$ and $r_{\rm sh}$ for $r_{\rm PNS}=20~{\rm km}$ are larger than those for $r_{\rm PNS}=10~{\rm km}$.
We confirm that our semi-analytic models are also consistent with these radii for $r_{\rm PNS} = 20~{\rm km}$ (see Section \ref{sec:rM_rsh}).

A larger $r_{\rm PNS}$ leads to a longer $t'_{\rm stall}$.
We derive the following functional form in the same manner as in Section \ref{sec:tstall}:
\begin{eqnarray}
    t'_{\rm stall}(\dot{M}_{\rm fb}, B_{\rm PNS})
    =
    1.5~[{\rm s}]~
    \left(\frac{\dot{M}_{\rm fb}}{{10^{-5}~\rm M_\odot~s^{-1}}}\right)^{-0.83}
    +1.2\times10^{1}~[{\rm s}]~
    \left(\frac{\dot{M}_{\rm fb}}{{10^{-5}~\rm M_\odot~s^{-1}}}\right)^{-1.3}
    \left(\frac{B_{\rm PNS}}{{10^{14}~\rm G}}\right)^{1.2},
    \label{eq:tstall_B1}
\end{eqnarray}
The accuracy of this fitting formula is better than $\approx 94\%$.

\subsection{Necessary condition for the submergence of the PNS magnetic field\label{sec:ineffective}}

We consider that the submergence of the PNS fields requires the abrupt increase in the gas density in the post-shock region.
The abrupt increase occurs when a sufficient amount of accreted matter accumulates at $r \sim r_{\rm M}$, and the gas density and temperature become high enough to induce rapid cooling (that is why the shock starts to stall at that time; see Section \ref{sec:time_evol}).
This may not be true when the supernova fallback ceases before the cooling becomes effective.
As we have ignored the duration of the supernova fallback in the above analysis, we investigate this effect.

To investigate the impact of the finite duration of the fallback, we perform a test simulation based on our fiducial model, {\tt B14Mm5R10}.
The setup of the test simulation is as follows.
As an initial condition, we take the physical quantities from the fiducial model at $t \approx 0.07~{\rm s}$.
We note $t_{\rm stall}\approx 0.25~{\rm s}$ in the fiducial model (see the dashed line in Figure \ref{fig:figure4}).
By setting $\rho=\rho_{\rm fl}$ and $p=p_{\rm fl}$ at $r=r_{\rm out}$, we effectively stop the fallback accretion at $t\sim 0.1~{\rm s}$.
In ghost cells outside $r=r_{\rm out}$, we impose $u^r=\max(u^r,0)$ and adopt the zero-gradient boundary condition for $\rho$ and $p$.

\begin{figure*}[tb]
\centering
\includegraphics[width=\linewidth]{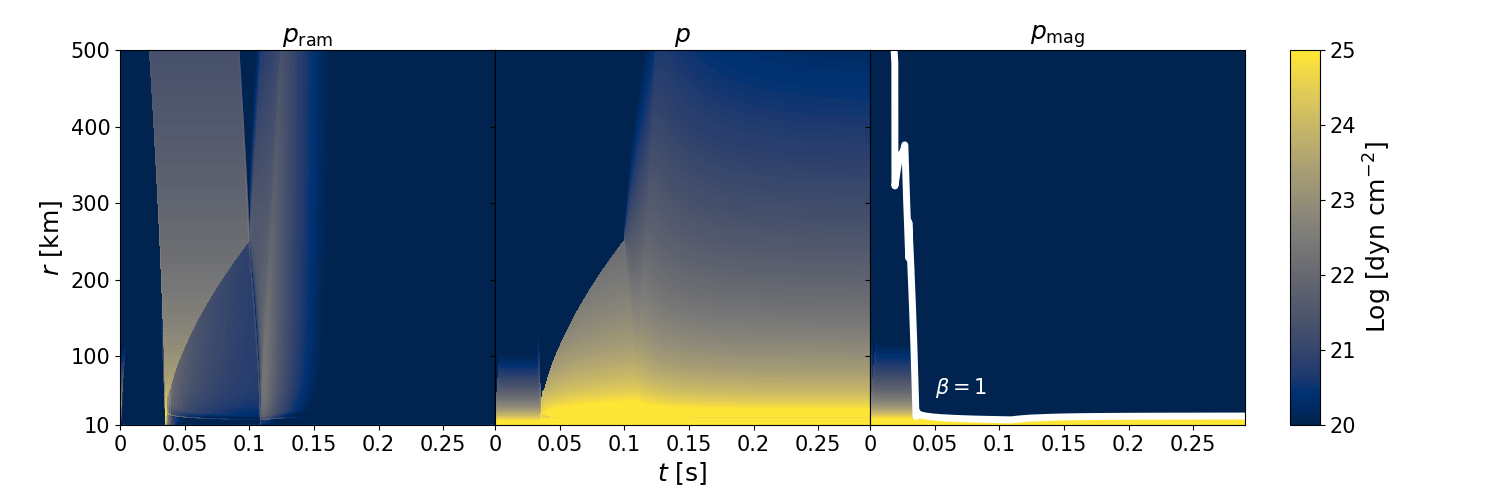}
\caption{
Time-sequenced images illustrating the radial profiles of the three different pressures in a test simulation.
In this test simulation, the initial conditions are extracted from the physical quantities at $t \approx 0.07~{\rm s}$ in model {\tt B14Mm5R10}.  
By adopting $\rho=\rho_{\rm fl}$ and $p=p_{\rm fl}$ at $r=r_{\rm out}$, we effectively stop the fallback accretion at $t \sim 0.1~{\rm s}$ (see Section \ref{sec:ineffective} for details).  
All labels are the same as in Figure \ref{fig:figure4}.
\label{fig:figure6}}
\end{figure*}

Figure \ref{fig:figure6} displays time-sequenced images of radial profiles of the three different pressures.
Due to the absence of the free-fall region, the accreted material redistributes itself to maintain the hydrostatic equilibrium.
As a result, the post-shock gas expands to reduce its density and temperature (see the middle panel).
The simulation does not show the abrupt increase in the density, suggesting that the PNS magnetosphere will not be submerged for $t'_{\rm stall}>t_{\rm fb}$.
Therefore, the condition $t'_{\rm stall} < t_{\rm fb}$ needs to be satisfied for the submergence.
The subsequent evolution of the hydrostatic atmosphere may become important for the submergence that occurs at a stage much later than the PNS phase.
We will discuss the late-time submergence in Section \ref{sec:limitation}.

{\section{Discussion}\label{sec:discussion}}

\subsection{Implications for the Diversity in Young isolated NS} \label{sec:implications}

We propose possible solutions for the supernova fallback based on the following two conditions and discuss the diversity of young isolated NSs.
The first condition is a necessary condition for the submergence of the PNS field (Section \ref{sec:ineffective}).
The second condition is related to an effect of the PNS rotation, which is ignored in our numerical models.
The rotating PNS can drive relativistic outflows and repel the fallback matter \citep[e.g.,][]{Zhong2021}.
We take this effect into account based on the theoretical study of \citet[][]{Zhong2021}.

We have suggested that the inequality $t'_{\rm stall} < t_{\rm fb}$ needs to be satisfied for the submergence of the PNS field (see Section \ref{sec:ineffective}).
Using Equations (\ref{eq:tstall_B}) and (\ref{eq:tstall_B1}), we can rewrite the relation $t'_{\rm stall}<t_{\rm fb}$ as
\begin{eqnarray}
    9.8\times10^{-2}~[{\rm s}]~
    \left(\frac{\dot{M}_{\rm fb}}{{10^{-5}~\rm M_\odot~s^{-1}}}\right)^{-0.69}
    +
    2.0\times10^{-1}~[{\rm s}]~
    \left(\frac{\dot{M}_{\rm fb}}{{10^{-5}~\rm M_\odot~s^{-1}}}\right)^{-1.2}
    \left(\frac{B_{\rm PNS}}{{10^{14}~\rm G}}\right)^{1.3}
    &<&t_{\rm fb},
    \label{eq:tstall_lt_tfb}\\
    1.5~[{\rm s}]~
    \left(\frac{\dot{M}_{\rm fb}}{{10^{-5}~\rm M_\odot~s^{-1}}}\right)^{-0.83}
    +1.2\times10^{1}~[{\rm s}]~
    \left(\frac{\dot{M}_{\rm fb}}{{10^{-5}~\rm M_\odot~s^{-1}}}\right)^{-1.3}
    \left(\frac{B_{\rm PNS}}{{10^{14}~\rm G}}\right)^{1.2}
    &<&t_{\rm fb},
    \label{eq:tstall_lt_tfb1}
\end{eqnarray}
for $r_{\rm PNS}=10~{\rm km}$ and $r_{\rm PNS}=20~{\rm km}$, respectively.

The condition related to the PNS rotation is as follows \citep[][]{Zhong2021}.
The spin-down luminosity is expressed as \citep[][]{Spitkovsky2006}
\begin{eqnarray}
    L_{\rm spin}=\frac{B_{\rm PNS}^2r_{\rm PNS}^6\Omega_{\rm PNS}^4}{4}(1+\sin^2\chi),
\end{eqnarray}
where $\Omega_{\rm PNS}=2\pi/P$ is the angular velocity of the PNS, $P$ is the spin period, and $\chi$ is the offset angle between the rotation and dipole axes.
We take $\chi=0$ for simplicity.
\citet{Zhong2021} assumed that the relativistic outflows and fallback matter typically encounters at the fallback radius, $r_{\rm fb}\sim (M_{\rm PNS}t_{\rm fb}^2)^{1/3}$.
From the condition of $L_{\rm spin}<\dot{M}_{\rm fb}v_{\rm ff}^2(r_{\rm fb})$, we obtain \citep[see also][for more generalized case]{Shigeyama2018}
\begin{eqnarray}
  \dot{M}_{\rm fb}>\frac{1}{2}L_{\rm spin}M_{\rm PNS}^{-2/3}t_{\rm fb}^{2/3}.
  \label{eq:spin_outflow}
\end{eqnarray}
Since the spin-down time scale, $\sim 10~{\rm yr}$, is much longer than $t_{\rm fb}$, the spin-down luminosity $L_{\rm spin}$ is regarded as a constant with time.
We employ $P = 0.02~{\rm s}$ as a fiducial value in Equation (\ref{eq:spin_outflow}), which is shorter than the pulse periods observed in CCOs \citep[$\sim0.1~{\rm s}$,][]{Enoto2019,Borghese2023}.

\begin{figure*}[tb]
\centering
\includegraphics[width=\linewidth]{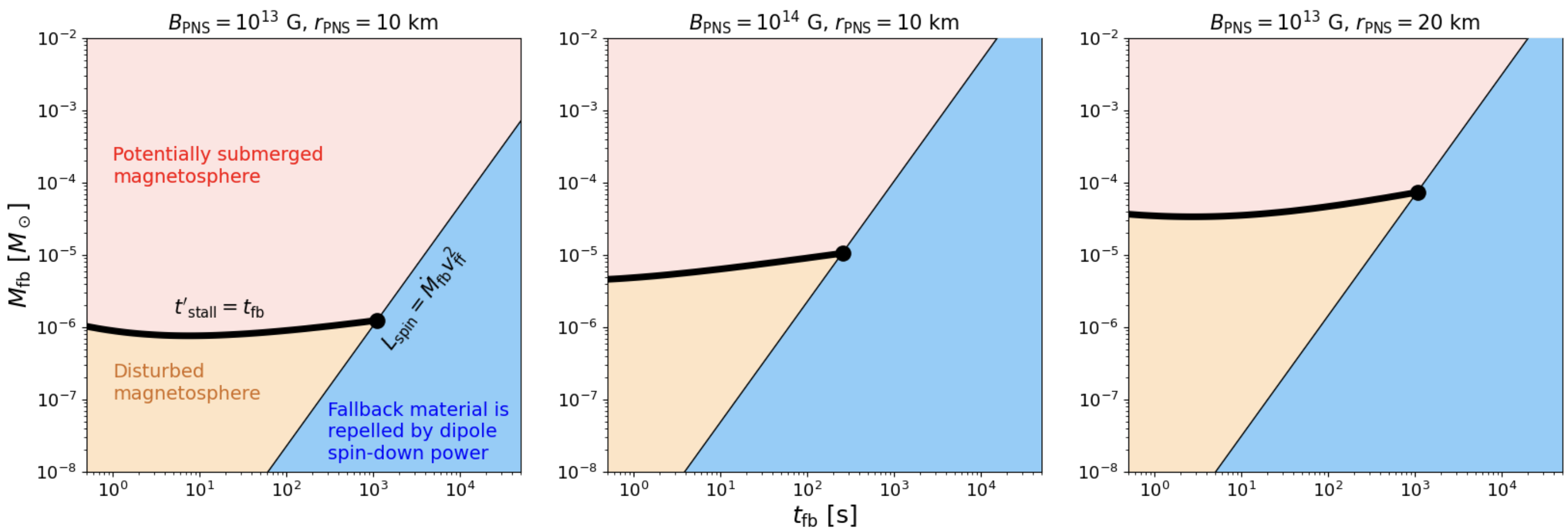}
\caption{
Possible solutions of the supernova fallback illustrated as a function of fallback mass and time.
We employ $P=0.02~{\rm s}$.
We show three cases of 
$(B_{\rm PNS},r_{\rm PNS})=(10^{13}~{\rm G},10~{\rm km})$ (left), 
$(B_{\rm PNS},r_{\rm PNS})=(10^{14}~{\rm G},10~{\rm km})$ (middle), and 
$(B_{\rm PNS},r_{\rm PNS})=(10^{13}~{\rm G},20~{\rm km})$ (right).
The thick and thin lines stand for $t'_{\rm stall}=t_{\rm fb}$ (Equation \ref{eq:tstall_lt_tfb}) and $L_{\rm spin}=\dot{M}_{\rm fb}v^2_{\rm ff}$ (Equation \ref{eq:spin_outflow}), respectively.
The fallback material can submerge the PNS magnetosphere for $t'_{\rm stall}<t_{\rm fb}$ and $L_{\rm spin}<\dot{M}_{\rm fb}v_{\rm ff}^2$ (red-shaded region).
The fallback material is repelled by the spin-down power for $L_{\rm spin}>\dot{M}_{\rm fb}v_{\rm ff}^2$ (blue-shaded region).
The fallback material disturbs the PNS magnetosphere for $t'_{\rm stall}>t_{\rm fb}$ and $L_{\rm spin}<\dot{M}_{\rm fb}v_{\rm ff}^2$ (orange-shaded region).
We represent the critical point branching into the three different solutions as a black point.
We consider that CCOs form in the red-shaded region, whereas the other types of NSs are born in the orange- and/or blue-shaded regions.
\label{fig:figure7}}
\end{figure*}

Figure \ref{fig:figure7} displays the possible solutions.
We have used the initial value of the accretion rate for the fallback matter, $\dot{M}_{\rm fb}=(2/3)(M_{\rm fb}/t_{\rm fb})$ (see Equation \ref{eq:fallback}).
Three cases are presented: 
$(B_{\rm PNS},r_{\rm PNS})=(10^{13}~{\rm G},10~{\rm km})$ (left), 
$(B_{\rm PNS},r_{\rm PNS})=(10^{14}~{\rm G},10~{\rm km})$ (middle), and 
$(B_{\rm PNS},r_{\rm PNS})=(10^{13}~{\rm G},20~{\rm km})$ (right).
The thick and thin lines represent $t'_{\rm stall}=t_{\rm fb}$ and $L_{\rm spin}=\dot{M}_{\rm fb}v^2_{\rm ff}$, respectively.
In the region where $t'_{\rm stall}<t_{\rm fb}$ and $L_{\rm spin}<\dot{M}_{\rm fb}v_{\rm ff}^2$ (red-shaded region), the supernova fallback can form a new crust and submerge the PNS magnetosphere.
In the region where $L_{\rm spin}>\dot{M}_{\rm fb}v_{\rm ff}^2$ (blue-shaded region), the fallback material is repelled by the dipole spin-down power and therefore does not interact with the PNS magnetosphere.
In this case, the NS magnetic field is determined solely by internal amplifications during the core-collapse and PNS phases.
In the region where $t'_{\rm stall}>t_{\rm fb}$ and $L_{\rm spin}<\dot{M}_{\rm fb}v_{\rm ff}^2$ (orange-shaded region), the fallback accretion is halted before the PNS magnetosphere is submerged, allowing the magnetosphere to survive.
During the fallback phase, the accreting material interacts with the PNS and disturbs its magnetosphere.
We speculate that the disturbance leads to an external amplification of the magnetic field.
We will discuss the amplification mechanisms induced by the supernova fallback later.

Figure \ref{fig:figure7} shows that the critical points (black points), which branch into three different solutions, are located around 
$(t_{\rm fb},M_{\rm fb})\approx(10^3~{\rm s},10^{-6}~{\rm M}_\odot)$ in the left panel,
$(t_{\rm fb},M_{\rm fb})\approx(3\times10^2~{\rm s},10^{-5}~{\rm M}_\odot)$ in the middle panel,
and
$(t_{\rm fb},M_{\rm fb})\approx(10^3~{\rm s},7\times10^{-5}~{\rm M}_\odot)$ in the right panel.
From the left and middle panels, the red-shaded region becomes narrower for a stronger $B_{\rm PNS}$ within the plotted parameter range.
In addition, the $M_{\rm fb}$ required for the submergence increases as $r_{\rm PNS}$ increases (see left and right panels).

The solutions in Figure \ref{fig:figure7} will provide insight into the diversity of the young isolated NSs.
The CCOs can form in the red-shaded region, while the other types of the NSs are born in the orange- and/or blue-shaded regions.
The solutions mean that when $B_{\rm PNS}=10^{13}~{\rm G}$ and $P=0.02~{\rm s}$, the NS population can be dominated by the CCOs within the typical ranges of $M_{\rm fb}=10^{-(4-1)}~{\rm M}_\odot$ and $t_{\rm fb}=10^{0-3}~{\rm s}$.
This may suggest that the CCOs are common in the young isolated NSs \citep{deLuca2008,Torres2016}.
In fact, despite the small number of the known CCOs, observational studies implied that the birth rate of the CCOs is as large as the other types of the NSs \citep{Gaensler2000}.

Figure \ref{fig:figure7} allows us to assess whether the necessary condition derived in this study is consistent with previous studies.
Here, we refer to \citet{Ho2011} and \citet{Vigano2012} for comparison.
Their approach differs from ours in that they focused on the long-term evolution of crustal magnetic fields (i.e., submerged fields), rather than on neutrino cooling in the fallback flows.
They solved the induction equation and obtained the re-emerged magnetic field strength at the NS surface as a function of time.
By comparing their result with CCO observations, they inferred a fallback mass of $M_{\rm fb}\gtrsim 10^{-3}~{\rm M}_\odot$ for $(B_{\rm PNS},r_{\rm PNS})\approx(10^{14}~{\rm G},10~{\rm km})$.
The middle panel of Figure \ref{fig:figure7} shows that their inferred value exceeds $M_{\rm fb}$ on the $t'_{\rm stall}=t_{\rm fb}$ line.
Thus, the necessary condition is not contradict with their models.


The supernova fallback may lead to an external amplification of the PNS magnetic field.
When the fallback matter reaches the PNS without submerging the surface magnetic fields (orange-shaded region in Figure \ref{fig:figure7}), it will chaotically disturb the magnetosphere.
This chaotic process can lead to stronger and compressed multipolar magnetic fields \citep[see discussion in][]{Zhong2021}.
Furthermore, \citet{Barrere2022} proposed a magnetar formation scenario through the supernova fallback, in which the PNS field is amplified due to the Tayler-Spruit dynamo triggered by the PNS spin-up from the supernova fallback \citep[see also][]{Barrere2025}.
Although a detailed investigation of magnetic field evolution is beyond the scope of this study, we speculate that supernova fallback serves as an additional factor contributing to magnetic field amplification.

We discuss how the crystallization of the fallback material affects Figure \ref{fig:figure7}.
We have derived Equations (\ref{eq:tstall_lt_tfb}) and (\ref{eq:tstall_lt_tfb1}) by comparing $t_{\rm stall}'$ with $t_{\rm fb}$.
However, the submergence of the PNS field requires the new crust formation (see Figure \ref{fig:figure1}), which means that we need to take into account the time interval from when the shock stalls to when the gas density and temperature satisfy the crystallizing condition.
To estimate such a time interval, we have performed an additional simulation based on model {\tt BnMm2R10} (see Appendix \ref{sec:appendix1} for details).
The simulation shows that the time interval is of the same order as the shock stalling time scale.
Therefore, we conclude that the crystallization of the fallback material does not significantly affect Figure \ref{fig:figure7}.
The crystallization time scale may become longer if we work with a more complete and realistic EoS \citep{Suwa2014,Nakazato2025}, as the degenerate pressure suppresses the increase in the gas density.
We will discuss this point in Section \ref{sec:limitation}.

\subsection{Model Limitations}\label{sec:limitation}

Regarding Figure \ref{fig:figure7}, we discuss possible multidimensional effects, which are ignored in our simulations.
Recent multidimensional simulations of the supernova explosion suggested that the fallback gas may have a large angular momentum \citep[see][and references therein]{Janka2022}.
When such material accretes onto the PNS, an accretion disk will form around the PNS magnetosphere \citep{Takahashi2017,Inoue2023,Inoue2024}.
The differential rotation of the disk opens up the PNS dipole field, which enhances $L_{\rm spin}$ \citep{Parfrey2017}.
Furthermore, MHD instabilities may also play an important role.
The surface at $r_{\rm M}$, where the magnetic pressure balances the radiation pressure, can be unstable to the magnetic Rayleigh-Taylor instability \citep[see e.g.,][]{Cumming2001,Stone2007}.
This instability leads to dynamic mixing around $r \sim r_{\rm M}$, suppressing the submergence of the PNS magnetosphere \citep[][]{Bernal2013}.
Therefore, the $\dot{M}_{\rm fb}$ required for the submergence will increase.

Realistic geometries of the dipole magnetic field and accretion flows can also lead to multidimensional effects.
The large-scale field geometry and magnetic tension, which are ignored in the present study, can form accretion mountains around the magnetic poles \citep{Melatos2001,Payne2004,Payne2007,Mukherjee2017}.
As the mountains spread, they will compress the PNS magnetic fields toward the equator \citep{Fujisawa2022, Yeole2025}, thereby enhancing the local magnetic pressure and suppressing the submergence in this region.
Furthermore, multidimensional simulations of core-collapse supernova showed that neutrino-driven wind and accretion occur at the same time \citep{Bollig2021,Nakamura2025,Vartanyan2025}, implying the presence of anisotropic fallback accretion flows.
Such anisotropic flows may partially submerge the dipole field, forming an NS with complex small-scale multipole fields \citep{Payne2004,Suvorov2020}.
When the NS spins down via multipole radiation, its spin-down rate is smaller than that of dipole radiation at a fixed magnetic field strength.
Therefore, the NS will be identified as the CCO if its spin-down is interpreted as originating from the dipole radiation.
The partial submergence can be achieved with a fallback rate smaller than that derived in this study.

The long-term evolution of the accreted material remains unexplored due to the limited duration of our simulations.
This study has focused on the submergence during the PNS phase, in which the remnant formed by core collapse (i.e., the PNS) contracts via neutrino cooling.
This phase lasts for several tens of seconds after the core bounce.
However, a submergence may also occur in a stage much later than the PNS phase.
The post-shock gas redistributes itself to maintain hydrostatic equilibrium after the fallback phase for $t_{\rm stall}'>t_{\rm fb}$ (see, Section \ref{sec:ineffective}).
The hydrostatic atmosphere will then cool via neutrino emission \citep{Combi2025} and/or radiation \citep{Takahashi2018,Abarca2018} on a time scale much longer than the duration of the PNS phase.
The atmosphere may submerge the magnetic fields at the crustal level in the post-PNS phase.
The necessary condition derived in this study does not account for such late-time submergence.
Therefore, we cannot rule out the possibility that the late-time submergence may occur even if the necessary condition is not satisfied.
To clarify the scenario, simulations that trace the long-term evolution of the atmosphere are required.

Finally, we make a few remarks about caveats for the EoS adopted in this study.
We assume a specific heat ratio of $\Gamma=4/3$ for the EoS and calculate the temperature from $T=(3p/a_{\rm rad})^{1/4}$ \citep{Chevalier1989}.
However, this is still an approximation.
To assess the validity of this approximation, we perform an additional simulation for $r_{\rm NS}=10~{\rm km}$ using the Helmholtz EoS \citep{Timmes2000} (see Appendix \ref{sec:degenerate} for details).
The simulation shows that the pair pressure increases the fluid pressure by up to a factor of two, resulting in an increase in the shock radius by a similar factor \citep{Houck1991}.
The resulting shock stalling time scale is longer than that in the original case with $r_{\rm NS}=10~{\rm km}$, but shorter than that in the original case with $r_{\rm NS}=20~{\rm km}$.
Therefore, the realistic $t_{\rm stall}'=t_{\rm fb}$ line for $r_{\rm NS}=10~{\rm km}$ will lie between those for $r_{\rm NS}=10~{\rm km}$ and for $r_{\rm NS}=20~{\rm km}$ in Figure \ref{fig:figure7}.
In addition, the additional simulation shows that while the electron degeneracy pressure is dynamically negligible for $t\le t_{\rm stall}$, it becomes non-negligible for $t>t_{\rm stall}$ \citep{Bernal2010}.
Therefore, the pressure will increase the crystallization time scale.
Since the qualitative behavior and overall trends remain unchanged, we retain the ideal-gas approximation for the remaining parameter survey.
To summarize this section, the solutions in Figure \ref{fig:figure7} require further updates.
However, we believe that our results will offer valuable guidance in investigating the diversity of young isolated NSs through multidimensional simulations.

{\section{Summary} \label{sec:conclusion}}

To understand the observational diversity of the young isolated NSs, we performed one-dimensional GRMHD simulations of supernova fallback onto a magnetized PNS, while taking neutrino cooling into account.
We demonstrated that the PNS magnetosphere, the post-shock region, and free-fall region appear (Figure \ref{fig:figure3}).
The accretion shock expands in an early stage, whereas it stalls and recedes in a later stage due to the neutrino cooling (Figure \ref{fig:figure4}).
Based on our numerical models, we derived a necessary condition for the submergence of the PNS magnetic field.
We summarize our findings in the following.

The one-dimensional simulations enabled us to study the time evolution of the fallback material over wide parameter ranges including the fallback mass accretion rate ($\dot{M}_{\rm fb}=10^{-(5-2)}~{\rm M}_\odot~{\rm s}^{-1}$), the magnetic field strength at the PNS surface ($B_{\rm PNS}=10^{13-15}~{\rm G}$, and $B_{\rm PNS}=0$ for comparison), and the PNS radius ($r_{\rm PNS}=10$ and $20~{\rm km}$).
From these simulations, we revealed that an abrupt increase in the gas density occurs after the shock stalls because of the neutrino cooling.
The abrupt increase potentially leads to the formation of a new crust that confines the PNS magnetosphere.

We developed semi-analytic models that are consistent with the magnetospheric and shock radii ($r_{\rm M}$ and $r_{\rm sh}$, respectively) when the shock stalls (Section \ref{sec:rM_rsh}).
The semi-analytic models are based on the following two relations: (1) the pressure balance between the magnetic pressure of the PNS field and the radiation pressure in the post-shock region (Equation \ref{eq:semi_ana1}), and (2) the balance between the release rate of the gravitational energy and energy loss rate due to neutrino cooling at $r=r_{\rm M}$ (Equation \ref{eq:semi_ana2}).
Therefore, both $r_{\rm M}$ and $r_{\rm sh}$ decrease with $\dot{M}_{\rm fb}$ and increase with $B_{\rm PNS}$ (see top and middle panels of Figure \ref{fig:figure5}).
Our semi-analytic model reduces to the analytical solution in \citet{Chevalier1989} when $B_{\rm PNS}=0$.

We found that the shock stalling time scale can be regarded as the waiting time for the new crust formation (Section \ref{sec:tstall}).
When the accretion rate is high ($\dot{M}_{\rm fb}=10^{-(3-2)}~{\rm M_\odot~s^{-1}}$), the shock stalling time scale is nearly equal to the time scale of the neutrino cooling.
When the accretion rate is low ($\dot{M}_{\rm fb}=10^{-(5-4)}~{\rm M_\odot~s^{-1}}$), the shock stalling time scale consists of the cooling time scale and the time delay for neutrino cooling to become effective.
Thus, the shock stalling time scale is long for a low $\dot{M}_{\rm fb}$ or a strong $B_{\rm PNS}$ (see the bottom panel of Figure \ref{fig:figure5}).

The dynamics of the supernova fallback depends on $r_{\rm PNS}$ (Section \ref{sec:dependence_on_rPNS}).
Since the gravitational acceleration at the PNS surface is smaller for larger $r_{\rm PNS}$, the radiation pressure and the cooling rate are reduced.
Therefore, both $r_{\rm M}$ and $r_{\rm sh}$ are larger for larger $r_{\rm PNS}$.
Furthermore, due to the lower cooling rate, the shock stalling time scale becomes longer for larger $r_{\rm PNS}$.

We proposed a necessary condition for the submergence of the PNS magnetosphere (Section \ref{sec:ineffective}).
The submergence requires the abrupt increase in the gas density in the post-shock region during the fallback phase.
Therefore, the shock stalling time scale needs to be shorter than the fallback time scale for the submergence.

Based on the necessary condition, we updated possible solutions of the supernova fallback to better understand the origin of the diversity of NS magnetic fields (Figure \ref{fig:figure7} in Section \ref{sec:implications}).
The solutions require further updates because they ignore some multidimensional effects and the electron-positron pair pressure.
However, we believe that the presented scheme will be a new benchmark for the scenarios of the NS diversity.

\begin{acknowledgments}

We thank Ryuichiro Akaho, Yudai Suwa, and Akira Dohi for fruitful discussions.
This work was supported by JSPS KAKENHI grant Nos. 
JP24KJ0143, JP25K17439 (A.I.), JP22K14074, JP22KK0043, JP21H04487 (S.T.), JP24K00668, JP23H04899, JP22H00130 (K.K.), and JP24K00672, JP21H04488, JP24K00678 (H.R.T).
A part of this research has been funded by the MEXT as "Program for Promoting Researches on the Supercomputer Fugaku" 
(Toward a unified view of the universe: from large-scale structures to  planets, JPMXP1020200109; A.I. and H.R.T.).
Numerical computations were performed with computational resources provided by Cray XC 50 at the Center for Computational Astrophysics (CfCA) of the National Astronomical Observatory of Japan (NAOJ), the FUJITSU Supercomputer PRIMEHPC FX1000 and FUJITSU Server PRIMERGY GX2570 (Wisteria/BDEC-01) at the Information Technology Center, The University of Tokyo.

\end{acknowledgments}

\appendix

{\section{Estimation of the effects of neutrino scattering and absorption}
\label{sec:sca_abs}}

Our numerical models assumed that neutrino cooling operates in an optically thin regime (see Equation \ref{eq:optically_thin_cooling}).
To evaluate the validity of this approximation, we estimate the radii of the neutrinospheres for $\nu_{\rm e}$ and $\bar{\nu}_{\rm e}$, at which the neutrino depths $\tau_{\nu_{\rm e}}$ and $\tau_{\bar{\nu}_{\rm e}}$ become unity, respectively.

We consider neutrino scattering and absorption by free protons and neutrons.
We neglect the effect of neutrino-electron scattering because its cross section is much smaller than those of neutrino-proton and neutrino-neutron scatterings \citep{Shapiro1983}.
We write the electron, proton, and neutron masses as $m_{\rm e}$, $m_{\rm p}$, and $m_{\rm n}$, respectively ($m_{\rm p}\approx m_{\rm n}\gg m_{\rm e}$).
We define the cross sections for the absorption processes of $\bar{\nu}_e + p \rightarrow e^+ + n$ and $\nu_e + n \rightarrow e^- + p$ as 
\begin{eqnarray}
    \sigma_{\rm abs}(\bar{\nu}_{\rm e}p)
    &=&
    \sigma_0
    \left(\frac{1+3g_{\rm A}^2}{4}\right)
    \left(\frac{\varepsilon_{\bar{\nu}_{\rm e}}-\Delta_{\rm np}}{m_{\rm e}}\right)^2
    \left[1-\left(\frac{m_{\rm e}}{\varepsilon_{\bar{\nu}_{\rm e}}-\Delta_{\rm np}}\right)^2\right]^{1/2},\\
    \sigma_{\rm abs}(\nu_{\rm e}n)
    &=&
    \sigma_0
    \left(\frac{1+3g_{\rm A}^2}{4}\right)
    \left(\frac{\varepsilon_{\nu_{\rm e}}+\Delta_{\rm np}}{m_{\rm e}}\right)^2
    \left[1-\left(\frac{m_{\rm e}}{\varepsilon_{\nu_{\rm e}}+\Delta_{\rm np}}\right)^2\right]^{1/2},
\end{eqnarray}
respectively \citep[][]{Burrows2006}.
Here, $\sigma_0=1.705\times10^{-44}~{\rm cm^2}$, $g_{\rm A}$ is the axial–vector coupling constant ($g_{\rm A}\approx -1.23$), $\varepsilon_{\nu_{\rm e}}$ is the energy of $\nu_{\rm e}$, $\varepsilon_{\bar{\nu}_{\rm e}}$ is the energy of $\bar{\nu}_{\rm e}$, and $\Delta_{\rm np}=m_{\rm n}-m_{\rm p}$.
We assume that the correction for weak magnetism and recoil is negligible for simplicity.
We describe the cross sections in the elastic limit of neutrino-proton and neutrino-neutron scatterings as follows \citep[see e.g.,][]{Shapiro1983,Suwa2019}:
\begin{eqnarray}
    \sigma_{\rm sca, p}
    \sim \sigma_{\rm sca, n}
    = \frac{\sigma_0}{4}\left(\frac{\varepsilon_{\nu}}{m_{\rm e}}\right)^2.
\end{eqnarray}

The method for estimating the radii of the neutrinospheres is as follows.
We denote the number densities of electrons, protons, and neutrons as $n_{\rm e}$, $n_{\rm p}$, and $n_{\rm n}$, respectively.
We can then write the mass density as $\rho = m_{\rm p}(n_{\rm p}+n_{\rm n})$.
We assume the charge neutrality condition $n_{\rm e}=n_{\rm p}$.
We represent the electron fraction as $Y_{\rm e}=n_{\rm p}/(n_{\rm p}+n_{\rm n})$.
From these definitions, we get the following relations: $n_{\rm p}=\rho Y_{\rm e}/{m_{\rm p}}$ and $n_{\rm n}={(1-Y_{\rm e})}\rho/m_{\rm p}$.
The detailed estimation of the value of $Y_e$ is beyond the scope of this study, and we adopt $Y_{\rm e}=0.5$ as a fiducial value for simplicity \citep{Fryer2006}.
From the definitions explained above, we calculate the radii of neutrinospheres using the following equations:
\begin{eqnarray}
    \tau_{\nu_{\rm e}}(\varepsilon_{\nu_{\rm e}})
    &=&\int^\infty_r (
    \sigma_{\rm abs}(\nu_{\rm e}n){n_{\rm n}}
    +\sigma_{{\rm sca},{\rm p}}{n_{\rm p}}
    +\sigma_{{\rm sca},{\rm n}}{n_{\rm n}})dr,\\
    \tau_{\bar{\nu}_{\rm e}}(\varepsilon_{\bar{\nu}_{\rm e}})
    &=&\int^\infty_r (
    \sigma_{\rm abs}(\bar{\nu}_{\rm e}p){n_{\rm p}}
    +\sigma_{{\rm sca},{\rm p}}{n_{\rm p}}
    +\sigma_{{\rm sca},{\rm n}}{n_{\rm n}})dr.
\end{eqnarray}
Since the typical gas temperature at $r\sim r_{\rm M}$ is $T\sim 10^{11}~{\rm K}$, we calculate $\tau_{\nu_{\rm e}}$ and $\tau_{\bar{\nu}_{\rm e}}$ for $\varepsilon_{\nu_{\rm e}}=\varepsilon_{\bar{\nu}_{\rm e}}=10~{\rm MeV}$ \citep[see also][]{Akaho2024}.

In all models, when $t=t_{\rm stall}$, both $\tau_{\nu_{\rm e}}$ and $\tau_{\bar{\nu}_{\rm e}}$ are much smaller than unity even at $r=r_{\rm M}$, suggesting that neutrino scattering and absorption by free protons and neutrons can be neglected \citep[see][for more detailed estimation]{Sumiyoshi2012}.

{\section{Time scale for the crust formation}\label{sec:appendix1}}

Our simulations demonstrated that neutrino cooling effectively works on a time scale of $t'_{\rm stall}$, leading to an abrupt increase in the gas density at $r\approx r_{\rm M}$.
As a result, a new crust will form just above the PNS magnetosphere.
We aim to estimate the time scale for the crust formation.
We define the time scale as the period between $t_{\rm stall}$ and the time when the crystallization begins.

When Coulomb energy dominates the thermal energy in the accreted matter, ions will crystallize into a periodic lattice structure to form the NS crust.
The crystallization of ions occurs if the gas temperature cools down to a melting temperature $T_{\rm m}=(Z^2e_{\rm c}^2)(\Gamma_{\rm m}k)^{-1}(4\pi n_{\rm ion}/3)^{1/3}$ \citep[see e.g.,][]{Shapiro1983,Chamel2008}, where $Z$ is the typical proton number of the lattice component, $e_{\rm c}$ is the elementary charge, $\Gamma_{\rm m}$ is the ratio of Coulomb to thermal energies of the lattice at the melting point, $k$ is the Boltzmann constant, and $n_{\rm ion}$ is the number density of ion.
For simplicity, we assume that the crust is formed of a perfect crystal with a single nuclear species at lattice sites \citep{Chamel2008}.
Its mass number and mass fraction are denoted by $A=Z/Y_{\rm e}$ and $x_{\rm a}$, respectively.
We then write the number density of ions as $n_{\rm ion}=(\rho x_{\rm a})/(Am_{\rm p})$.
From these definitions, we can rewrite the melting temperature as \citep{Suwa2014}:
\begin{eqnarray}
    T_{\rm m}
    =
    5.6\times10^{9}~[{\rm K}]
    \left(\frac{\Gamma_{\rm m}}{175}\right)^{-1}
    \left(\frac{\rho}{10^{14}~{\rm g~cm^{-3}}}\right)^{1/3}
    \left(\frac{Y_{\rm e}}{0.5}\right)^{1/3}
    \left(\frac{x_{\rm a}}{0.1}\right)^{1/3}
    \left(\frac{Z}{28}\right)^{5/3}.
    \label{eq:crystal}
\end{eqnarray}
We adopt $\Gamma_{\rm m}=175$ and $x_{\rm a}=0.1$ as fiducial values \citep{Suwa2014}.
We consider nuclei with the magic numbers $Z=28$, $50$, and $82$ \citep{Haensel2001,Chamel2008}.

To estimate the time scale for crystallization, we continue the simulation in model {\tt BnMm2R10} until the temperature reaches the melting temperature defined by Equation (\ref{eq:crystal}).
The details of the setup are as follows.
To ensure sufficient spatial resolution while maintaining reasonable computational cost, we conduct the simulation in three successive stages.
We first set $r_{\rm out}$ to $20~{\rm km}$ with the number of numerical grid points of $N_r = 16384$.
Physical quantities at each grid point are initialized by linearly interpolating the data from model {\tt BnMm2} at time $t\approx 0.0354~{\rm s}$.
For grid points located inside the innermost radial point of model {\tt BnMm2}, we assign the values of the physical quantities at the innermost point of model {\tt BnMm2}.
We run a simulation by $t=0.0356~{\rm s}$, and then the resulting profile is used as the initial condition for the next stage. 
This procedure is repeated for $r_{\rm out}=10.1~{\rm km}$ and $10.01~{\rm km}$.
This approach ensures $\Delta r<H/10$.

\begin{figure}[tb]
\centering
\includegraphics[width=\linewidth]{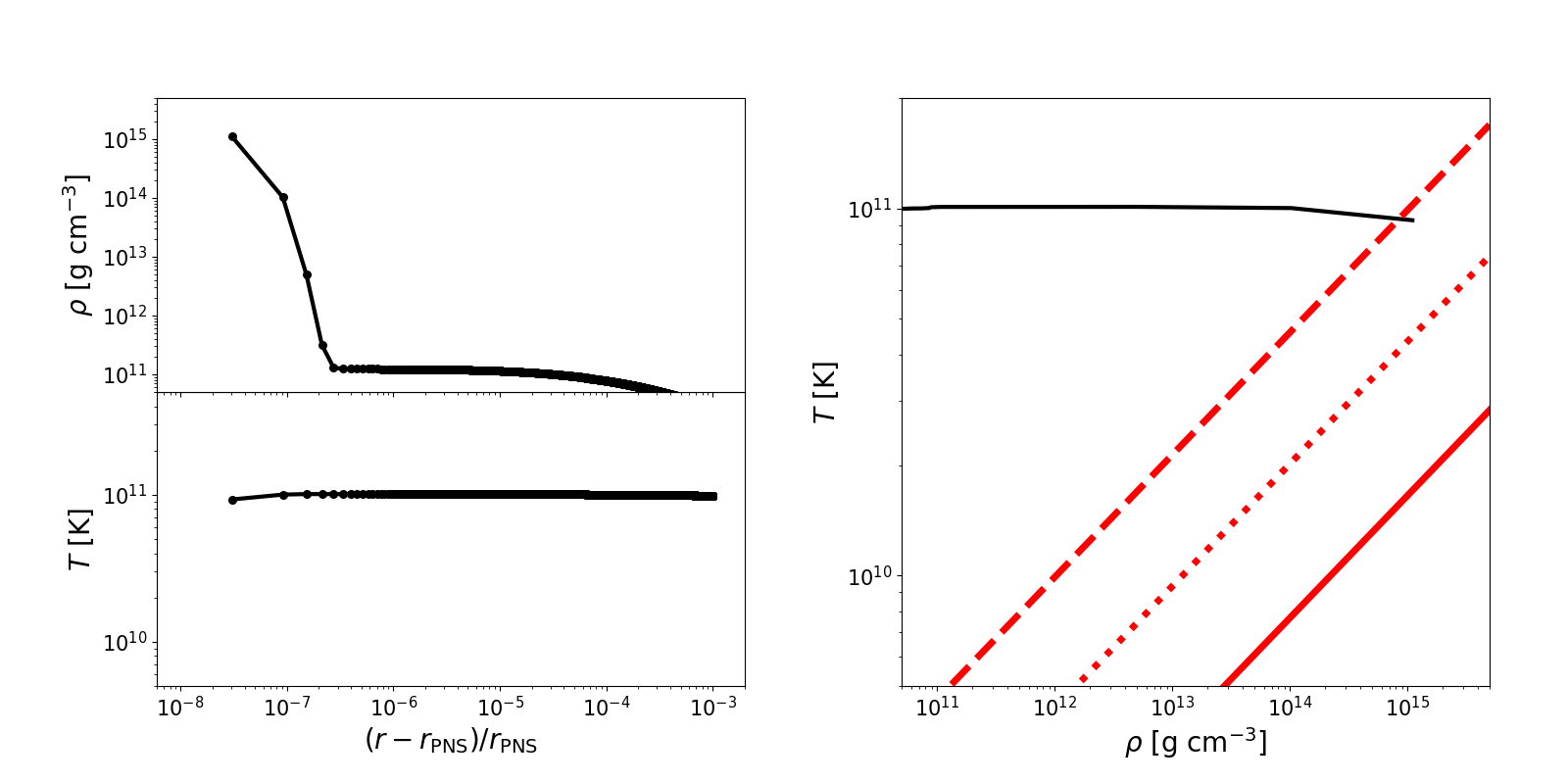}
\caption{
Left: Radial profiles of the gas density (top) and the gas temperature (bottom) in the vicinity of the PNS surface.
Right: Plots for the gas temperature against the gas density (black line).
Crystallizing temperature in the cases of $Z=28$ (solid), $Z=50$ (dotted), $Z=82$ (dashed) are represented by red lines (see Equation \ref{eq:crystal}).
All plots are measured at $t\approx 0.0357~{\rm s}$.
\label{fig:figure8}}
\end{figure}

The left panels of Figure \ref{fig:figure8} show the radial profiles of the gas density (top) and temperature (bottom) in the vicinity of the PNS surface at $t\approx 0.0357~{\rm s}$.
The gas density reaches $\sim 10^{15}~{\rm g~cm^{-3}}$ at the innermost grid, while the temperature remains nearly constant with radius.
The neutrino cooling via the URCA process efficiently works at the innermost grid due to its high density.

The right panel of Figure~\ref{fig:figure8} investigates the condition for the crystallization.
The black line displays the relation between $\rho$ and $T$ at $t\approx 0.0357~{\rm s}$, while the red lines represent $T_{\rm m}$ for $Z=28$ (solid), $Z=50$ (dotted), and $Z=82$ (dashed).
The figure indicates that the presented model satisfies the crystallization condition if $Z=82$.
We define $t_{\rm crust}$ as the time at which the model satisfies the condition for $Z=82$ ($t_{\rm crust}\approx0.0357~{\rm s}$).

The resulting time scale of the new crust formation is $t_{\rm crust}-t_{\rm stall}\sim 10^{-3}~{\rm s}$, which is roughly equal to $t_{\rm stall}'$.
Therefore, we conclude that the crystallization time scale does not significantly affect Figure \ref{fig:figure7} within the framework adopted in this study.

{\section{Results considering Helmholtz Equation of State}\label{sec:degenerate}}

We assumed an effective adiabatic index of $\Gamma=4/3$ for the EoS and calculated the temperature from the Stefan-Boltzmann law for simplicity.
To assess the impact of the EoS on our conclusions, we perform an additional test simulation employing a more realistic EoS.

We adopt the Helmholtz EoS \citep{Timmes2000} for the additional simulation, which is based on a tabulated interpolation of the Helmholtz free energy computed by the Timmes EoS \citep{Timmes1999}.
The Helmholtz EoS includes contributions from blackbody radiation, fully ionized nuclei, and degenerate and relativistic electrons and positrons.
The fluid pressure and its internal energy density are obtained by summing over these components.
When the temperature and density are specified, the Helmholtz EoS package returns the corresponding fluid pressure and internal energy density.

We describe the numerical methods and setup below.
The total pressure is used as a primitive variable in the default version of the {\tt UWABAMI} code, whereas we instead use the temperature in this simulation.
We adopt the recovery scheme for the primitive variables proposed by \citet{Noble2006} \citep[see also][]{Siegel2018}.
The fluid is assumed to consist of free protons, neutrons, and electrons, with $Y_{\rm e}=0.5$.
The PNS radius is set to $r_{\rm PNS}=10~{\rm km}$.
We investigate the case with $B_{\rm PNS}=0$ and $\dot{M}_{\rm fb}=10^{-3}~{\rm M}\odot~{\rm s}^{-1}$.
We refer to this model as the Helmholtz model and compare it with models {\tt BnMm3R10} and {\tt BnMm3R20}.

\begin{figure}[tb]
\centering
\includegraphics[width=0.5\linewidth]{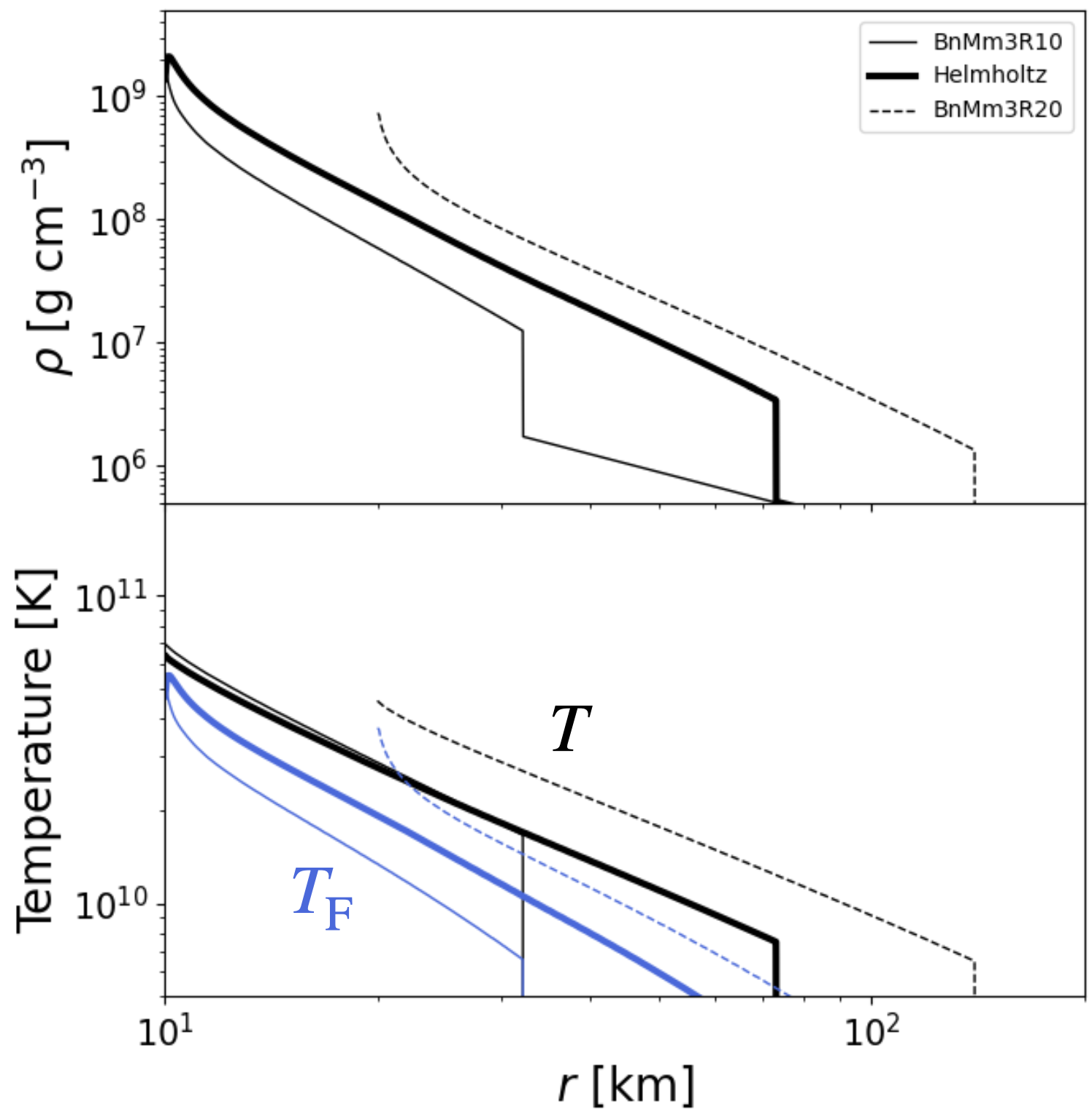}
\caption{
The radial profiles of the gas density (upper panel) and temperatures (lower panel) when the shock stalls.
The thick lines represent the results in the Helmholtz model, while the thin solid and the dashed lines indicate the results in models {\tt BnMm3R10} and {\tt BnMm3R20}, respectively.
The blue lines in the lower panel indicate the Fermi temperature.
\label{fig:figure9}}
\end{figure}

The upper panel of Figure \ref{fig:figure9} shows the radial profiles of the gas density when the shock stalls.
The thick line represents the result of the Helmholtz model, while the thin solid line and the dashed line indicate the results in models {\tt BnMm3R10} and {\tt BnMm3R20}, respectively.
In the post shock region, the resulting gas density in the Helmholtz model is greater than that in model {\tt BnMm3R10}.

The upper panel of Figure \ref{fig:figure9} demonstrates that the shock radius in the Helmholtz model is greater than that in the ideal case with  $r_{\rm NS}=10~{\rm km}$, but smaller than that in the ideal case with $r_{\rm NS}=20~{\rm km}$.
The pair pressure increases the fluid pressure by up to a factor of two, resulting in a larger shock radius \citep{Houck1991}.
The resulting shock stalling time scale is longer than that in the original case with $r_{\rm NS}=10~{\rm km}$, but shorter than that in the original case with $r_{\rm NS}=20~{\rm km}$.

The black lines in the lower panel of Figure \ref{fig:figure9} represent the radial profiles of the gas temperature.
The resulting temperature in the Helmholtz model is very close to that in the original case with $r_{\rm NS}=10~{\rm km}$ near the PNS surface.

We compare the resulting temperature with the Fermi temperature.
The blue lines in the lower panel of Figure \ref{fig:figure9} indicate the Fermi temperature.
We calculate the Fermi momentum from $p_{\rm F}=(3\pi^2 n_{\rm e})^{1/3}\hbar$.
The Fermi energy is given by $E_{\rm F}=\sqrt{(p_{\rm F}c)^2+(m_{\rm e}c^2)^2}-m_{\rm e}c^2$.
The Fermi temperature is described as $T_{\rm F}=E_{\rm F}/k$.
We find $T>T_{\rm F}$ for $t\le t_{\rm stall}$ but $T<T_{\rm F}$ for $t>t_{\rm stall}$.
Therefore, the degenerate pressure of the electrons is negligible for $t\le t_{\rm stall}$.
We note that the degenerate pressure cannot be neglected after the shock stalls.
Because the degenerate pressure suppresses the increase in the gas density, the crystallization time scale (see Appendix \ref{sec:appendix1}) will increase when we consider the Helmholtz EoS.
The relation $T<T_{\rm F}$ for $t>t_{\rm stall}$ is consistent with the result in \citet{Bernal2010}.

\bibliographystyle{aasjournal}
\bibliography{library}

\end{CJK*}
\end{document}